\documentclass[aps,rmp,showpacs,twocolumn,amsmath,amssymb,floatfix,superscriptaddress]{revtex4-1}

\usepackage{color}  
\usepackage{graphicx}
\usepackage{braket}
\usepackage{bbold}
\usepackage{lineno}
\begin{document}

\title[Author guidelines for IOP journals in  \LaTeXe]{Single atom detection in ultracold quantum gases: a review of current progress}

\author{Herwig Ott}

\address{Department of physics and Research Center OPTIMAS, University of Kaiserslautern, 67663 Kaiserslautern, Germany}

\begin{abstract}
The recent advances in single atom detection and manipulation in experiments with ultracold quantum gases are reviewed. The discussion starts with the basic principles of trapping, cooling and detecting single ions and atoms. The realization of single atom detection in ultracold quantum gases is presented in detail and the employed methods, which are based on light scattering, electron scattering, field ionization and direct neutral particle detection are discussed. The microscopic coherent manipulation of single atoms in a quantum gas is also covered. Various examples are given in order to highlight the power of these approaches to study many-body quantum systems.
\end{abstract}

\maketitle

\tableofcontents{}

\section{Introduction: Dealing with single particles}
Whenever new experimental techniques are developed, new physical observables become accessible. The detection of single particles is an outstanding example which has revolutionized our understanding of physics. This is true for the highest energy scales such as studied in particle accelerators as well as for the lowest energy scales, realized with ultracold quantum gases. In the realm of quantum optics and ultracold quantum gases, the detection of single particles goes hand in hand with superb preparation and manipulation techniques. The famous quote of Erwin Schr\"odinger "\textit{In the first place it is fair to state that we are not experimenting with single particles, any more than we can raise Ichthyosauria in the zoo}" \cite{Schrödinger1952} stands exemplary for the ever ongoing technological progress in physics.

While single particle detection was realized almost one hundred years ago by Charles Wilson with the help of his cloud chamber, the full potential of quantum research on the single particle level became accessible when experimentalists managed to control and read-out the internal and external degrees of freedom. This was first achieved for charged particles, starting with the study of single electrons in a Penning trap \cite{Wineland1973} and single ions in Paul traps \cite{Neuhauser1980,Wineland1981}. Meanwhile, the control over the motion and the internal states of trapped ions has provided unprecedented insight into the microscopic quantum world. The study of quantum jumps \cite{Sauter1986,Bergquist1986}, the observation of the quantum Zeno effect \cite{Itano1990}, the demonstration of quantum logical operations \cite{Monroe1995,SchmidtKaler2003} and a whole toolbox for quantum simulation with trapped ions \cite{Roos2012} have established cold ion systems as one of the leading platforms in quantum research. At the same time, the experimental capabilities to observe single particles have triggered new theoretical concepts like, e.g., the quantum trajectory approach \cite{Dalibard1992,Carmichael1993}.

The control of neutral particles is based on static magnetic fields and absorptive or dispersive light forces \cite{Phillips1998,Ketterle1999}. The magneto-optical trap (MOT) \cite{Raab1987}, a combination of a magnetic quadrupole field and six mutually orthogonal laser beams with properly chosen frequency and polarization, is the most convenient way to cool neutral particles. In a single-atom MOT \cite{Hu1994,Haubrich1996}, only one atom is present at the time and the continuous scattering of photons forms the basis of fluorescence imaging. By a careful analysis of the photon statistics, even higher atom numbers can be discriminated, culminating in the accurate counting of up to 1200 atoms \cite{Hume2013}. In a MOT the continuous absorption and emission of photons makes the coherent control of the atoms practically impossible, apart from very short timescales. This limitation has been overcome by the preparation of single atoms in optical dipole traps \cite{Schlosser2001,Kuhr2001,Alt2003,Schrader2004}, where photon scattering is highly suppressed. Such approaches are employed to work with few particles, studying for instance quantum walks \cite{Karski2009}, few-body quantum systems \cite{Serwane2011}, entanglement \cite{Wilk2010} or coherent few-body dynamics \cite{Barredo2015}. Experiments with single atoms interacting with a single mode radiation field \cite{Meschede1985,An1994,Nogues1999,Gleyzes2007} have also achieved an impressive level of control.

The advent of ultracold quantum gases has revolutionized the research field of many-body physics. The large number of available bosonic and fermionic atomic species, the variety of trapping potentials and the ability to tune the interaction offer a rich portfolio to study the ground state properties of many-body quantum systems and their dynamics. The realization of Bose-Einstein condensation in dilute atomic gases \cite{Anderson_1995,Bradley_1995,Davis_1995} has marked the beginning of a still ongoing expansion of experimental and theoretical activities \cite{Ketterle1999}. An important further milestone has been the implementation of strong interactions with help of so-called optical lattices - periodic potentials created by laser beams which mimic the potential landscape in a solid \cite{Bloch_2008}. This has led to the observation of the superfluid to Mott insulator transition in an ultracod gas \cite{Greiner2002}. Non-interacting and interacting fermionic quantum gases \cite{Ketterle2008} have also entered the stage and the powerful technique of Feshbach resonances \cite{Chin2010}, where the interaction strength between two atoms can be set by an external magnetic field, has enabled the observation of the BEC-BCS crossover in ultracold Fermi gases \cite{Randeria2014}. Further developments include the study of spinor systems \cite{Stamper-Kurn_2013} and artifical gauge fields \cite{Lin_2009, Lin_2011}. The conceptual simplicity of ultracold quantum gases, their purity and the high level of experimental control over their parameters make them ideal model systems to study fundamental questions and to benchmark theoretical calculations  \cite{Bloch2012}. Ultimately, they can help to tackle the most challenging problems in many-body physics such as high Tc superconductivity \cite{Dagotto2005}.

Many of the above listed many-body systems encode their properties in correlation functions and complex microscopic dynamics. A high resolution, single atom sensitive detection method gives access to many of these properties and greatly enlarges the prospects of ultracold quantum gas research. And there are more fields, where single atom detection and manipulation in an ultracold quantum gas bears great potential: (i) Transport of particles plays a central role in solid state systems. With the help of ultracold quantum gases, fundamental aspects of such transport processes can be studied under controlled conditions \cite{Brantut_2012,Ronzheimer_2013,Labouvie2015}. High resolution imaging techniques are thereby ideally suited to characterize the microscopic transport dynamics. (ii) Understanding the non-equilibrium dynamics of closed interacting quantum systems is another challenging problem in many-body physics~\cite{Polkovnikov_2011,Langen_2014,Eisert_2014}. The ability to design, control and measure tailored quantum systems on the single atom level paves the way towards understanding local and global properties of non-equilibrium dynamics. (iii) The evolution of an open many-body quantum system is governed by the coupling of the system with an environment. Engineering the action of the environment can be a tool to produce and stabilize interesting many-body quantum states \cite{Diehl_2008,Diehl_2011}. High resolution {\it in situ} addressability of a quantum gas is one way to realize this~\cite{Barontini_2013,Labouvie2015b}.

The many promising prospects of single atom detection in ultracold quantum gases are contrasted by the experimental difficulties of imaging many neutral atoms at small relative distances in an experimental setup with plenty of optics installed and limited spatial access. In the past years, several different experimental approaches have overcome these challenges. Experiments with metastable helium atoms \cite{Schellekens2005,Jeltes2007,Hodgman2011,Vassen2012} were among the first to detect single atoms from an ultracold atomic gas. The technique relies on the detection of the atoms in time of fight. After release from the trapping potential and a free fall of about half a meter, the metastable atoms hit a position sensitive detector and their three-dimensional density distribution is reconstructed. The single particle sensitivity has been exploited to study correlation functions of Bose-Einstein condensates, degenerate Fermi gases and thermal clouds. Later on, an optical analogue of this principle based on the fluorescence signal of atoms falling through an optical light sheet has been demonstrated \cite{Perrin2012}. The first optical \textit{in situ} imaging of single atoms in a three-dimensional optical lattice was realized in Ref.\,\cite{Nelson_2007} by means of optical fluorescence. With a lattice spacing of $5\,\mu$m, a high occupancy of excited states in each lattice site and an average filling of $1/2$, the system was not in the degenerate quantum regime.  \textit{In situ} detection of single atoms inside a quantum gas was first achieved by a scanning electron microscopy technique \cite{Gericke_2008}. Using a focused electron beam, atoms inside the quantum gas are ionized by electron impact ionization and detected. The spatial resolution of this technique is set by the diameter of the electron beam. To date, electron microscopy is still the imaging technique that achieves the highest resolution (about 150\,nm \cite{Gericke_2008}). Single atom-resolved optical fluorescence imaging of a quantum gas in a two-dimensional optical lattice has been achieved shortly afterwards for rubidium atoms \,\cite{Bakr_2009,Sherson_2010,Bakr2010} and later on for ytterbium atoms \cite{Miranda2015,Yamamoto2016}, lithium atoms \cite{Parsons2015,Omran2015} and potassium atoms \cite{Cheuk2015,Haller2015,Edge2015}. It relies on the spatially resolved detection of fluorescence light emitted by the atoms while they are exposed to light fields. During the exposure time the atoms remain pinned to the underlying lattice structure. This is achieved by either using laser cooling schemes such as optical molasses \cite{Bakr_2009,Sherson_2010,Yamamoto2016}, Raman sideband cooling \cite{Cheuk2015,Parsons2015,Omran2015} and so-called EIT cooling \cite{Haller2015,Edge2015} or by combining deep enough optical potentials for the ground and excited state of the atomic transition with a short exposure time \cite{Miranda2015}. These approaches are the only ones that combine \textit{in situ} imaging of atoms with single site resolution and almost unit detection fidelity. They have been used for a series of fundamental studies of strongly correlated quantum systems (see, for instance, Refs.\, \cite{Endres2011,Cheneau2012,Fukuhara2013,Preiss2015}). The successful imaging of lithium and potassium atoms represents the first single atom detection in fermionic quantum gases.

Detecting the position and the internal state of single atoms in a quantum gas is a powerful way to characterize a many-body quantum state. Manipulating single atoms -- coherently or incoherently -- is an add-on which draws a straight connection to non-equilibrium physics. This has been shown by dynamical studies of quantum gases subject to local excitations or density quenches \cite{Weitenberg2011,Endres2011,Fukuhara2013,Labouvie2015,Preiss2015}. Because the measurement of a particle has an influence on the remaining system, single atom detection in an ultracold quantum gas is also closely related to the study of open quantum systems \cite{Breuer2002}. Thereby, it literally provides a direct visualization of the quantum to classical transition.

The present review aims at a presentation of the state-of-the-art of single atom detection in ultracold quantum gases. The review is organized as follows: in Section II, we will briefly review the techniques employed for preparing and detecting single atoms in a more general context, including single ions, electrons and atoms. Many techniques for ultracold quantum gases are inspired or directly related to these experiments and many concepts can be clarified with the help of these few-body systems. Section III is the main part of this review and summarizes various techniques for single atom detection in a quantum gas based on ionization, neutral particle detection and photon scattering. The technical description of the detection is illustrated by experimental studies, which exemplify the potential of the imaging technique. Section IV is devoted to the manipulation of single atoms in a quantum gas. An outlook on the future perspectives and their impact on the research field of ultracold quantum gases is given in Section V.

\section{Preparation and detection of individual ions and atoms}
\label{sec:prep}

\subsection{Ions}

\begin{figure}[t!]
\begin{center}
\includegraphics[width=0.35\textwidth]{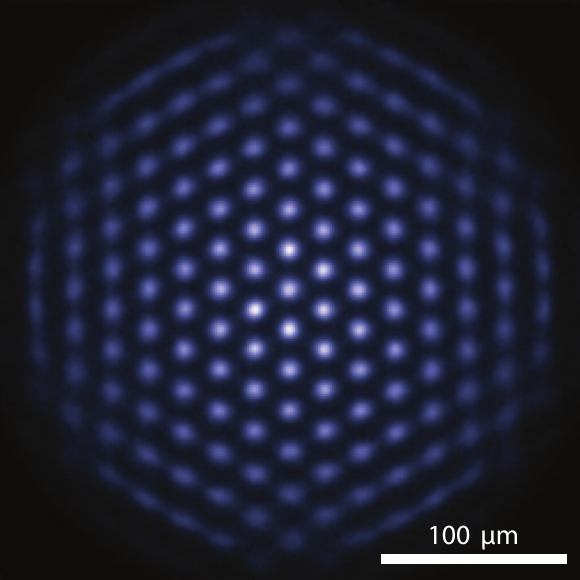}
\end{center}
\caption{Fluorescence image of a two-dimensional ion crystal consisting of several hundred beryllium ions in a Penning trap. The ions arrange in a triangluar lattice structure and the lattice constant is $d=20\,\mu$m. Employing spin-dependent dipole forces, an effective spin-spin interaction between the ions can be induced. Taken from Ref.\,\cite{Britton2012}}  \label{fig:ioncrystal}
\end{figure}

The preparation of ultracold ions relies on laser cooling techniques in a Penning or Paul trap. Penning traps employ a strong homogeneous magnetic field, which provides the confinement in one plane. An electrostatic field confines the ions in the perpendicular direction. Because they only use static fields, they are especially suited for precision measurements \cite{Brown1986,Blaum2010}. In the context of  quantum optics, many-body physics and quantum information processing, Paul traps are more common. They use time-dependent electric fields which are created by small electrodes of centimeter or millimeter size. The resulting trap geometries can be linear or planar. Alternatively, the electrodes can be implemented on a micro-fabricated chip, which allows for a small trap volume and large flexibility in the trap geometry. In most cases, the resulting confining potential for the ions can be approximated by a harmonic oscillator potential in all three directions (this also holds for the neutral atoms discussed in the next subsection)

\begin{equation}
V(x,y,z)=\frac{1}{2}m\omega_x^2x^2+\frac{1}{2}m\omega_y^2y^2+\frac{1}{2}m\omega_z^2z^2,
\label{eq:trappingpotential}
\end{equation}

where $\omega_i$ denotes the oscillation frequencies in the $i$-direction. The frequencies of an ion trap are in the kHz to MHz range and provide a steep and deep trapping potential which can exceed the thermal energy scale. Ions can then be trapped for months. The loading can be realized by ionizing neutral atoms from a background gas within the trapping volume. After the first demonstration of cooling and trapping a single Barium ion in a Paul trap \cite{Neuhauser1980} various laser cooling techniques have been developed for ions and we refer to the literature for a detailed discussion \cite{Stenholm1986,Diedrich1989,Lethokov1995,Adams1997,Eschner2003})
The fluorescence photons which are produced during the laser cooling allow for the detection of the ions by simply collecting them with a high numerical aperture imaging system. The photon detection is typically done with a sensitive CCD cameras. When more than one ion is trapped, the Coulomb repulsion between the ions leads to regular patterns such as linear chains, zig-zag structures or helical structures \cite{Birkl1992}. Fig.\,\ref{fig:ioncrystal} shows the fluorescence image of a trapped ion crystal in a planar Penning trap, showing Wigner crystallization \cite{Britton2012}. The typical distance between individually trapped ions is on the order of ten micrometer, which can be resolved by an imaging system with moderate resolution.

Apart from fluorescence imaging, ions can also be detected by a particle detector. Continuous or discrete dynode electron multipliers are used in many cases. Similar to photomultipliers and avalanche diodes, these detectors do not resolve the position of the particle, apart from specially fabricated detector arrays. The detection efficiency depends on the ion energy and can reach values close to 1 \cite{Koizumi2008}. Multi-channel plates (MCP) in combination with a phosphor screen plus CCD camera or in combination with a delay line anode provide a spatial resolution down to 100\,$\mu$m. MCPs have a lower detection efficiency of 40-60\%, which originates from the ratio between the plain surface and the holes.

\onecolumngrid
\begin{center}
\begin{figure*}[h]

\includegraphics[width=0.8\textwidth]{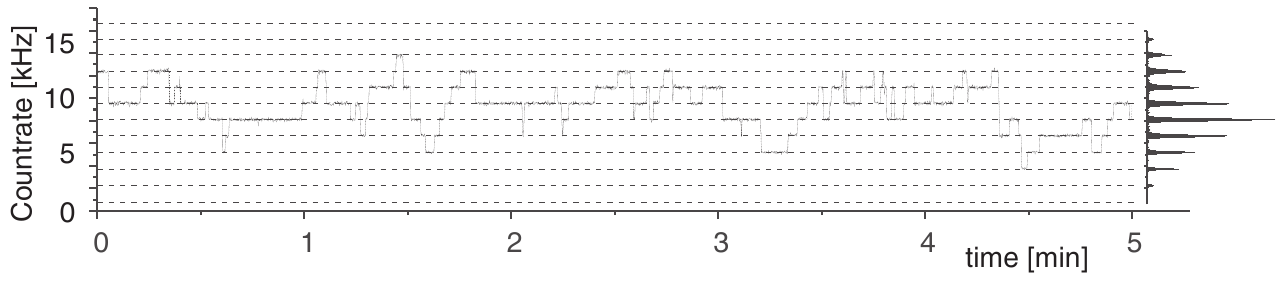}

\caption{Fluorescence signal of individual cesium atoms in a magneto-optical trap. On a timescale of several tens of seconds, the atoms enter or leave the trap. The fluoresecence detection is sensitive enough to distinguish the different atom numbers. The histogram shows the time-averaged atom number distribution. Taken from Ref.\,\cite{Gomer2001}}  \label{fig:MOT}
\end{figure*}
\end{center}
\twocolumngrid

\subsection{Neutral atoms}

The forces which can be exerted on neutral atoms by magnetic and electric fields are much weaker than the electrostatic forces in an ion trap. The potential energy of an atom with a magnetic moment of one Bohr magneton in a magnetic field of 1000\,G is five orders of magnitude smaller than that of a singly charged ion in an electric potential of 1\,V. For the coupling of the electric dipole moment of an atom to the oscillating electric field of a laser, similar arguments apply. Traps for neutral atoms are therefore much shallower compared to ion traps and the oscillation frequencies are in the Hz and kHz regime. This goes along with a much smaller trap depth. Laser cooling of neutral atoms therefore requires more sophisticated techniques and the single particle detection is more challenging. The magneto-optical trap (MOT) \cite{Raab1987} is the first stage in practically all ultracold atom experiments and was the first system, where a single neutral particle has been trapped \cite{Hu1994,Haubrich1996}. Even though a MOT features a rather large trap depth compared to conservative magnetic and optical traps, the energy transfer in a collision with a background gas atom suffices to remove the atom from the trap. This limits the lifetime of a neutral particle in a MOT to a few tens of seconds, depending on the vacuum level \cite{Cohen2011}. 

In a single-atom MOT, the continuous laser cooling leads to a stochastic motion of the atom which performs a random walk in an area of several tens of micrometers. At the same time, the laser cooled atom scatters photons, which are collected by a detector. Fig.\,\ref{fig:MOT} shows the fluorescence signal of a few atom MOT over several minutes \cite{Gomer2001}. During that time, the atoms stochastically enter and leave the trapping area. In order to efficiently confine the trapped atom and to keep the overall atom number in the MOT small, the magnetic field gradient $b$ of the MOT is one order of magnitude higher compared to a standard MOT with large atom number. The large magnetic field gradient decreases the trapping volume, whose radius scales as $b^{-14/3}$ \cite{Haubrich1993}. By a careful analysis of the fluorescence statistics, up to 1200 atoms can be counted accurately \cite{Hume2013}. Such capabilities are a powerful feature of magneto-optical traps and are often combined with the previous manipulation in conservative magnetic and optical traps. Single atom counting in a MOT has also applications in atom trap trace analysis of elements with extremely low abundance \cite{Jiang2011,Ritterbusch2014}.

In a MOT, the trapping principle requires the continuous absorption and emission of photons. Coherent manipulation of atoms is therefore practically impossible, apart from very short timescales. The coherent manipulation of atoms and further applications in quantum information processing require the loading of single atoms in optical trapping potentials \cite{Grimm2000}. In such a trap, photon scattering is strongly suppressed because the dispersive atom-light interaction which is responsible for the conservative trapping potential scales as $I/\Delta$ ($I$ is the laser intensity and $\Delta$ is the detuning from the atomic resonance), while the scattering rate, which is responsible for the heating, scales as $I/\Delta^2$. For a large detuning $\Delta$ the latter can be made sufficiently small and the atoms can be trapped for several tens of seconds without significant heating. 

The most simple optical dipole trap for single atoms is a tightly focused laser beam which intersects a magneto-optical trap \cite{Frese2000}. Due to the small focus (typically ten micrometer or less), the trap depth is in the mK regime and exceeds the temperature of the atoms in the MOT by at least one order of magnitude. The atoms can therefore be trapped by simply switching off the MOT at the end of the loading sequence. The number of atoms which is loaded in a dipole trap usually follows a Poissonian distribution \cite{Fuhrmanek2010}. However, for sufficiently strong confinement (one micrometer focus diameter), the trap volume is so small that the simultaneous presence of two or more atoms in the trap leads to light-assisted collisions. Such a collision leads to pairwise loss of the atoms. As a consequence, only no atom or one atom are present in the trap and the number statistics is strongly sub-Poissonian with a probability of about 50\,\% for the presence of a single atom \cite{Schlosser2001,Weber2006}. As only a few milliwatt of laser power are needed for such small traps, arrays of microtraps can be created with spatial light modulators \cite{Knoernschild2010,Nogrette2014}. A near deterministic loading scheme which achieves an efficiency of 87\,\% has been reported in Ref.\,\cite{Grünzweig2010}.

One-dimensional optical lattices are also used to trap single atoms loaded from a MOT \cite{Kuhr2001,Alt2003}. The optical lattice provides a strong confinement in one direction and can be used to trap several atoms at well defined sites in the lattice \cite{Schrader2004,Miroshnychenko2006}. Optical cavities offer another way to study single atoms. Due to the cavity, the interaction between the light and the atom is strongly increased. The small spacing between the two cavity mirrors (usually below one millimeter) prevents direct loading and the atoms are often launched from a MOT into the cavity \cite{Ye1999,Hood2000,McKeever2003,Puppe2007}.

After being loaded into the dipole trap, the temperature \footnote{The temperature of a single trapped particle is usually defined as the average over the occupied energy states.} of the atoms is given by the temperature of the MOT. In some cases, the temperature is additionally lowered by sub Doppler cooling mechanism like optical molasses or polarization gradient cooling \cite{Adams1997}. In order to image the atoms in the dipole trap, the MOT and the dipole trap are operated simultaneously. The fluorescence is captured by a high NA objective and imaged on a sensitive CCD camera. The fluorescence light is typically collected for tens or hundreds of milliseconds during which $10^3$ to $10^4$ photons per atom are detected. As the atoms are hold in place by the dipole trap, a fluorescence image of the atom is obtained.

\begin{figure}[t!]
\begin{center}
\includegraphics[width=0.3\textwidth]{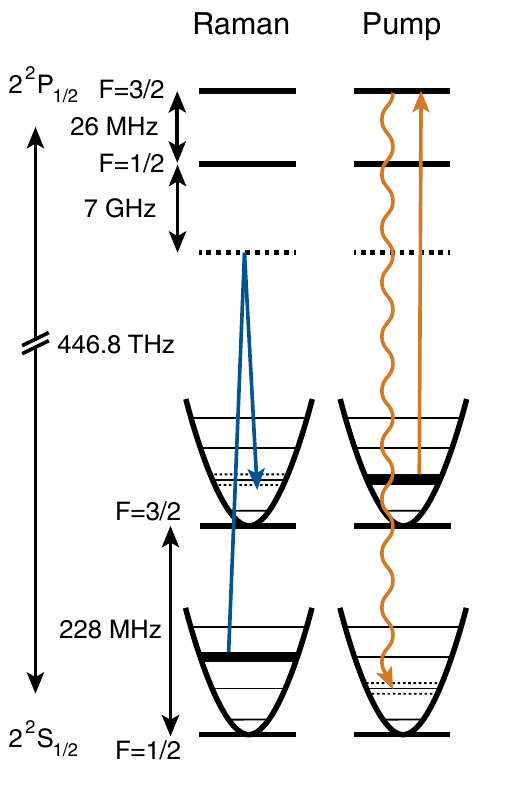}
\end{center}
\caption{Raman sideband cooling for lithium 6 as realized in a quantum gas microscope. The Raman transition from the $F=1/2$ to the $F=3/2$ hyperfine state of lithium removes one vibrational quantum from the particle's energy (left). Within the repumping process (right) the vibrational state is not altered. Taken from Ref.\,\cite{Parsons2015}.}  \label{fig:Raman}
\end{figure}

For many applications, the temperature of the atoms and the corresponding thermal motion in the dipole trap are not relevant. However, especially in the context of many-body quantum system, the full control over the external degree of freedom is necessary. Two strategies are pursued to achieve this. The first one seeks to directly cool the atoms to the ground state of the trapping potential. Raman sideband cooling is the most commonly used technique. It has been explored for trapped ions in great detail \cite{Eschner2003} and has been transferred to atoms trapped in optical lattices \cite{Vuletic1998,Belmechri2013,Patil2014,Cheuk2015,Parsons2015,Omran2015}, optical dipole traps \cite{Lester2014} and optical cavities \cite{Boozer2006,Reiserer2013}. Fig.\,\ref{fig:Raman} sketches the basic principle. Raman sideband cooling is most efficiently done in the so-called Lamb-Dicke regime, where the photon recoil energy is much smaller than the harmonic oscillator energy $\hbar \omega$. This ensures that the internal state of the atom is only slightly coupled to its motional state and the latter is practically conserved in a spontaneous emission event. The tight confinement in a single atom trap fulfills in general this condition. Because spontaneous emission is part of the cooling cycle, it can be used for fluorescence imaging. Raman sideband cooling is particularly useful when optical molasses do not achieve a low enough temperature to keep the atoms pinned to the trap.

The second strategy starts from a Mott insulator state in an optical lattice, which features ground state occupation from the very beginning. The desired atomic distribution is obtained by removing all other atoms. This approach bears the advantage that a large number of atoms can be prepared simultaneously. A more detailed discussion and examples for this are given in sections III.B and IV.

\section{Detecting single atoms in a quantum gas}

\label{sec:singleatomdetection}

The interest in ultracold quantum gases ranges from the quantum simulation of many-body systems, quantum optics, quantum information processing, transport phenomena and open system control to sensing and metrology. Many of these topics can benefit from single atom detection capabilities. While single atom sensitivity is mandatory for tasks like quantum information processing \cite{Raussendorf_2001} or the measurement of higher order correlations and string order \cite{Endres2011,Preiss2015}, also more ''classical'' problems such as the precise measurement of density distributions \cite{Wurtz_PRL_2009,Vogler2013}, pair correlation measurements \cite{Guarrera_correlations_2011} or impurity physics \cite{Spethmann2012} can take advantage from it. 

The requirements for single atom detection in a quantum gas are more stringent compared to ion systems: the interesting length scales are smaller and the trapping potential for neutral atoms is much shallower. This becomes particularly clear in the case of the Hubbard- and Bose-Hubbard Hamiltonian, which are extensively studied in the research field of ultracold quantum gases \cite{Bloch2005,Bloch2012}: 

\begin{eqnarray}\label{eq:BHM}
H_{\mathrm{BH}}&=&-J\sum_{\left< i,j\right> }b_i^\dagger b_j +\frac{U}{2}\sum_i n_i(n_i-1) \\
H_{\mathrm{H}}&=&-J\sum_{\left< i,j\right>,\sigma} c_{i,\sigma}^\dagger c_{j,\sigma} +U\sum_i n_{i\uparrow}n_{i\downarrow}.
\end{eqnarray}

Here, $J$ is the tunneling coupling, $U$ is the onsite interaction energy, and $\left< i,j \right>$ denotes nearest neighboring lattice sites. The bosonic creation and annihilation operators are $b_i^\dagger$ and $b_i$ ($n_i=b_i^\dagger b_i$), the fermionic creation and anihilation operators are $c_{i,\sigma}^\dagger$ and $c_{i,\sigma}$, where $\sigma=\uparrow$, $\downarrow$ denotes the spin state and $n_{i\uparrow}=c_{i\uparrow}^\dagger c_{i\uparrow}$, $n_{i\downarrow}=c_{i\downarrow}^\dagger c_{i\downarrow}$.

In a typical experimental realization of these Hamiltonians with ultracold atoms, the wavelength of the fluorescence light is comparable to the lattice spacing. As a consequence, the distance between two sites in an optical lattice is on the order of the optical diffraction limit of the detection method. Therefore, a single site resolved imaging technique has to come close to this limit. Also in bulk systems, where no lattice potential is present, various length scales are below one micrometer. For instance, the healing length in a Bose-Einstein condensate \cite{Ketterle1999} typically amounts to several hundreds of nanometers and the size of a vortex \cite{Madison2000} or a soliton \cite{Becker2008} is directly related to it. At densities above $10^{12}$\,cm$^{-3}$, the average interatomic distance is also below one micrometer. Another example is the spatial extension of the pair correlations in a strongly interacting one-dimensional quantum gas which is given by the interparticle distance \cite{Guarrera_2012}. Sub-micrometer resolution is therefore a necessity for efficient single atom detection in a quantum gas. 

The production of ultracold quantum gases is a subject on its own and not part of this review. Detailed descriptions can be found in \cite{Ketterle1999,Ketterle2008,Cohen2011}. We here assume that the quantum gas is already prepared and resides in an optical dipole trap \cite{Grimm2000} or a magnetic trap \cite{Fortagh2007}. In the following, we discuss single atom detection based on direct neutral particle detection, ionization, and light scattering. 

\subsection{Direct particle detection}

\begin{figure}[t!]
\begin{center}
\includegraphics[width=0.5\textwidth]{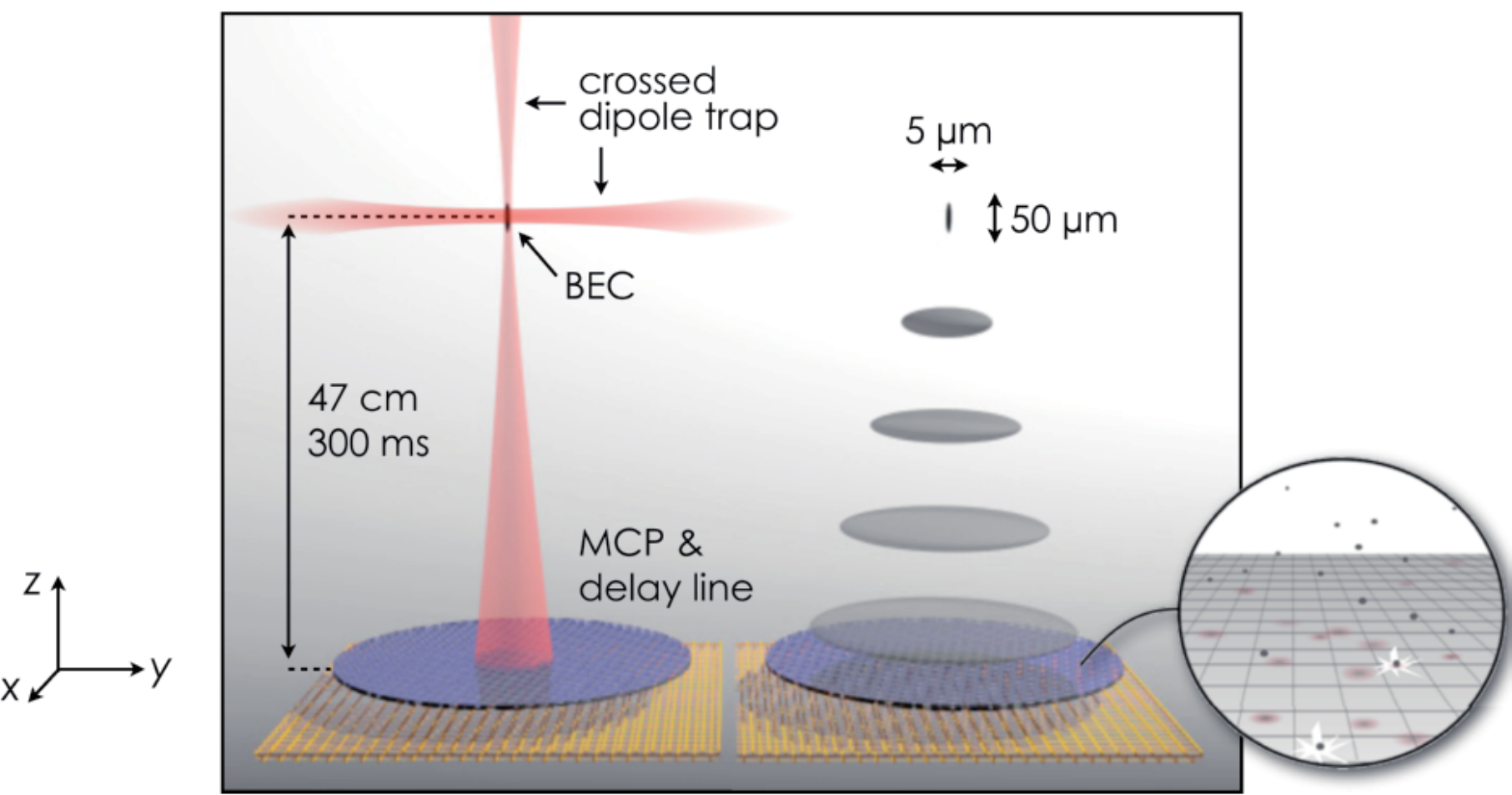}
\end{center}
\caption{Schematics for single particle detection of metastable noble gas atoms. The atoms are prepared in a spin triplet state and fall onto a micro channel plate, once the trapping potential is switched off. The internal energy of the atoms creates free electrons from the surface, which are subsequently detected by the channel plate. With permission from C. Westbrook.}  \label{fig:MCP_Helium}
\end{figure}

A neutral particle in a metastable internal state can store enough energy to release electrons upon contact with a surface. This principle is exploited in ultracold atom experiments with metastable noble gas atoms \cite{Vassen2012}. The atoms are prepared in the spin triplet state, which also provides a magnetic moment to trap the atoms magnetically. Experiments in the degenerate regime have been carried out with bosonic $^4$He and fermionic $^3$He atoms \cite{Santos2001,Schellekens2005,Tychkov2006,Hodgman2011}. The detection principle is sketched in Fig.\,\ref{fig:MCP_Helium}. After release from the trap, the atoms hit a microchannel plate which is located below the cloud. Once the atoms touch the surface of the MCP, the internal state is quenched and the release energy ejects electrons from the surface. These electrons are accelerated in the MCP, leading to a detectable signal. Due to the spatial resolution in two dimensions and the additional temporal information in the third direction, this technique provides the reconstruction of the full three-dimensional atom distribution. The technique works only in a time of flight arrangement and reaches a detection efficiency of about 25\,\% This is lower than the typical detection efficiency of an MCP for charged particles (40 - 60\,\%) and has its origin in the different detection process. The single particle sensitivity has enabled the observation of a series of fundamental quantum optical effects, ranging from Hanbury Brown and Twiss correlations for bosons \cite{Schellekens2005} and fermions \cite{Jeltes2007}, four wave mixing \cite{Perrin2007}, third order correlation functions \cite{Hodgman2011} to the demonstration of a Hong-Ou Mandel interferometer for atom pairs \cite{Lopes2015}. 

A related optical fluorescence technique follows a similar working principle and measures the transit of single atoms through a light sheet which is located below the atomic sample. While the atoms are falling through the light sheet, a CCD camera records the fluorescence traces. This has been used to measure Hanbury Brown and Twiss correlations across the Bose-Einstein condensation threshold \cite{Perrin2012}. The technique can in principle be adopted to all atomic species used in cold atom experiments.

\subsection{Ionization}

Ionization of a neutral particle and subsequent ion detection is another technique that can be used to detect single particles in ultracold quantum gases. It also allows for time-resolved studies of the quantum gas. The ionization process can be performed by electron impact ionization, photo- and field ionization as well as intrinsic ionizing collisions. The detection of the ions is realized with a channeltron, a discrete electron multiplier or a multi-channel plate.

\subsubsection{Electron impact ionization}
\begin{center}
\begin{figure*}[h]

\includegraphics[width=0.32\textwidth]{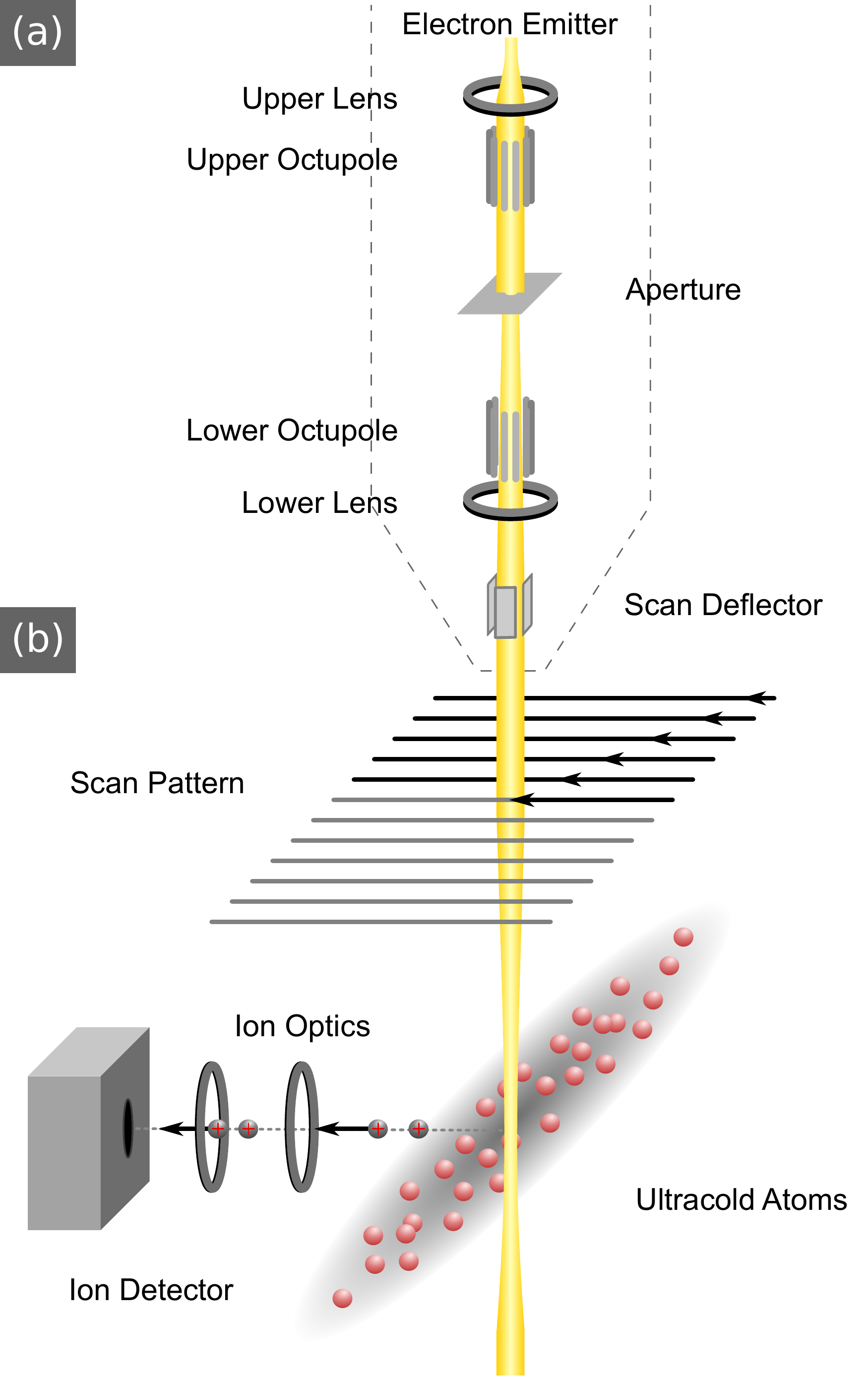}
\includegraphics[width=0.32\textwidth]{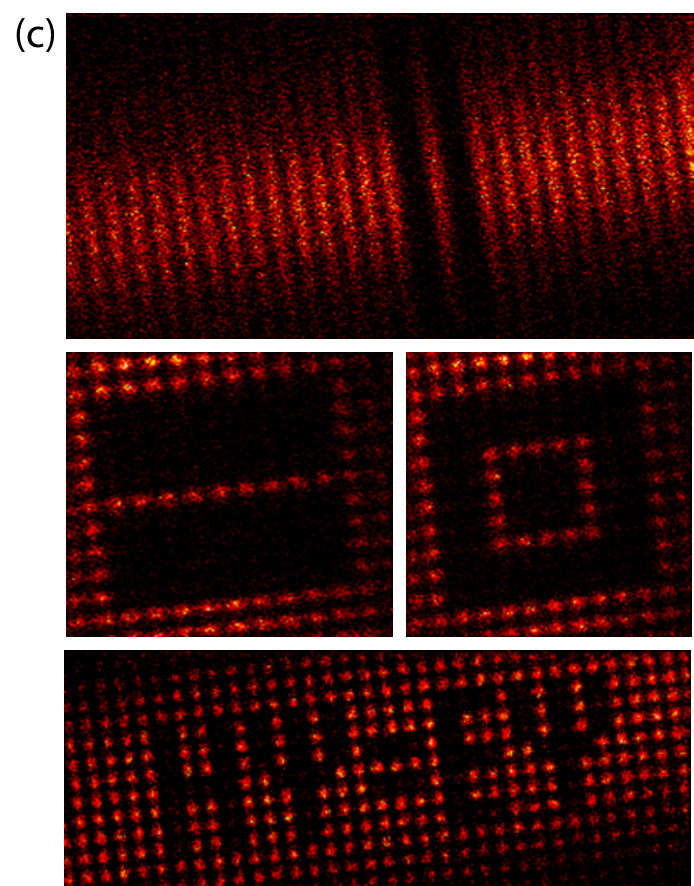}

\caption{Scanning electron microscopy of ultracold quantum gases. (a) The electron column, represented by the dashed line, provides a focused electron beam, which intersects the atomic sample, prepared in an optical dipole trap.  (b) The electron beam is scanned across the cloud. Electron impact ionization produces ions, which are guided with an ion optical system towards a channeltron detector. Taken from Ref.\,\cite{Wurtz_applPhysB2010}. (c) Preparation and detection of arbitrary patterns of Bose-Einstein condensed atoms in a two-dimensional optical lattice with a spacing of 600 nm. Taken from Ref.~\cite{Wurtz_PRL_2009}.}  \label{fig:ebeam}
\end{figure*}
\end{center}

The first high resolution imaging of single atoms in a quantum gas was realized with a scanning electron microscope \cite{Gericke_2008}. This approach employs the electron impact ionization of neutral atoms with the help of a focused electron beam \cite{Gericke_2006}. The working principle is the following (see Fig.\,\ref{fig:ebeam}): an electron column provides a focused electron beam which is scanned across an ultracold quantum gas~\cite{Gericke_2007}. The diameter of the electron beam is between 100-500\,nm and the beam current ranges from 10-500\,nA. Electron impact ionization creates ions, which are extracted with an electrostatic field and detected by a channeltron. The small diameter of the electron beam ensures a high spatial resolution, whereas the ion detection provides single-atom sensitivity. A typical imaging sequence consists of a rectangular scan pattern of 100\,ms duration, in which a few hundred atoms are detected. The overall detection efficiency is limited by the branching ratio between electron impact ionization and non-ionizing collisions and amounts to 10 - 20\,\%. As the cross-section for electron-atom scattering ($\sigma_{tot} = 1.78\pm 0.14 \times 10^{-16}$ cm$^{2}$~\cite{Wurtz_2010}) is eight orders of magnitude smaller than the absorption cross-section of a resonant photon, the atomic cloud is optically thin for the electron beam. For typical parameters, only one out of 500,000 incident electrons undergoes a collision.

The high spatial resolution of the imaging technique is illustrated in Fig.\,\ref{fig:ebeam}c. The images show a Bose-Einstein condensate of rubidium atoms which is loaded in a one- or two-dimensional optical lattice. From a quantitative evaluation one can deduce a spatial resolution better tahn 150\,nm~\cite{Gericke_2008}. The technique can also be used for single-site manipulation in an optical lattice. To this purpose, atoms are removed selectively from individual sites by means of collisions with the focused electron beam. In this way, arbitrary patterns of occupied lattice sites can be produced~\cite{Wurtz_PRL_2009}. Several examples are presented in Fig.~\ref{fig:ebeam}. A more detailed description of scanning electron microscopy of ultracold gases and more experiments exploiting this technique can be found in Ref.\,\cite{Santra2015}.

\onecolumngrid
\twocolumngrid

\subsubsection{Photoionization}
A scanning probe technique can also be realized with the help of photoionization, where ground state atoms are ionized in a single- or multiple photon absorption process \cite{Anderlini2004,Kraft2007,Viteau2010,Stibor2010}. For most atoms, single photon ionization is challenging, as the required wavelength is in the ultraviolet spectral region. In the case of rubidium - the workhorse of cold atom experiments - the wavelength is 297\,nm. While this can in principle be achieved with the help of frequency doubling \cite{Manthey_2014}, two-photon \cite{Courtade2004} or even three-photon \cite{Dodhy1987} excitation schemes are more common because of the larger cross sections. The ionization cross section for single photon ionization from the ground state of rubidium is $\sigma_{\mathrm{5s}}=1.7\times10^{-20}$\,cm$^{2}$ \cite{Lowell2002}. The cross section from the first excited state of rubidium is already a factor of 1000 larger, $\sigma_{\mathrm{5p}}=1.48\times10^{-17}$\,cm$^{2}$ \cite{Gabbanini1997}. In order to achieve a high spatial resolution that is competitive with other techniques, the focus of the laser beams has to be smaller than 1\,$\mu$m. The high intensity that is created by the small focus relaxes the conditions for beam power and a few milliwatt of laser power leads to an ionization rate in the MHz range, even for a single photon process. Care has to be taken due to the dispersive light forces, which can attract or repel the remaining atoms. Note that the ionization in a multi-photon process can be made state-selective. Photoionization has been applied to thermal rubidium atoms \cite{Kraft2007} and Bose-Einstein condensates of rubidium \cite{Viteau2010}.

\subsubsection{Field ionization}
Ground state atoms need electric fields of $10^{7}-10^{8}$\,V/cm to be field ionized. Sharp tips such as that of carbon nanotubes are capable to provide electric fields of this magnitude. In Ref.\,\cite{Grüner2009} it has been shown that carbon nanotubes can rapidly field ionize rubidium ground state atoms in the vicinity of the tip. This has the potential to establish a spatially resolved atom detection close to surfaces.

In the research field of Rydberg physics, field ionization is the workhorse for atom detection. Electric fields on the order of $1-2$ kV/cm are already sufficient to instantly ionize the highly excited atoms. With the help of an inhomogeneous electric field close to a tip, spatial correlations between Rydberg atoms have been detected \cite{Schwarzkopf2011,Schwarzkopf2013}. The sensitivity of Rydberg atoms to focused electron beams has also been studied \cite{Manthey_2014}. With the help of single site single atom fluorescence imaging (see also next subsection), individual Rydberg atoms in an optical lattice have been detected and an emerging crystalline structure, which is mediated by the long range interaction between the Rydberg atoms, has been observed \cite{Schauss2015}. While the field of Rydberg physics has strong connections to the research of ultracold quantum gases, a more detailed discussion of the detection mechanisms is beyond the scope of this review. We refer to Refs.\,\cite{Gallagher1988,Saffman2010,Marcassa2014} for a deeper insight into this field.

\subsection{Light scattering}

We begin this section with a brief reminder of the basic fluorescence and absorption processes. We approximate the atom with an ideal two-level system with transition frequency $\omega_0$ and decay rate $\gamma$. The atom is driven by a light field $E(t)=E_0\sin \omega_L t$, where $\omega_L$ is the angular frequency of the light field and $E_0$ is its amplitude. The photon scattering rate of an atom exposed to the light field is then given by

\begin{equation}\label{eq:scatteringrate}
\Gamma_\mathrm{scat}=\frac{\gamma}{2}\frac{\Omega^2/2}{\Omega^2/2+\gamma^2/4+\Delta^2},
\end{equation} 

where $\Omega=d E_0/\hbar$ is the Rabi frequency, $d$ is the dipole matrix element of the transition and $\Delta$ is the detuning. The scattered photons are emitted into the full solid angle and their spatially resolved detection forms the basis for fluorescence imaging. In absorption imaging, the attenuation of the light beam during the propagation through the cloud is measured. This follows Beer's law and yields

\begin{equation}\label{eq:absorption}
I(x,y)=I_0(x,y)\mathrm{Exp}\left[-\sigma\int  n(x,y,z)dz\right],
\end{equation}

where $n(x,y,z)$ is the density of target atoms, $I_0(x,y)$ is the initial intensity and $I(x,y)$ is the intensity behind the atomic ensemble. The absorption cross section $\sigma$ is connected to the scattering rate (\ref{eq:scatteringrate}):

\begin{equation}
\sigma=\frac{\hbar\omega_L}{2I_{\mathrm{sat}}}\frac{\gamma^2}{1+4\Delta^2/\gamma^2}.
\end{equation} 

Here, $I_\mathrm{sat}=\pi h c \gamma /(3\lambda^3)$ is the saturation intensity of the transition \cite{Cohen1998} and we have assumed that $I_0<I_\mathrm{sat}$. A more detailed description of atom-light interaction can be found, e.g., in \cite{Cohen1998,Aspect2010,Cohen2011}.

\subsubsection{Fluorescence in an optical lattice}

\begin{figure}[t!]
\begin{center}
\includegraphics[width=0.85\textwidth]{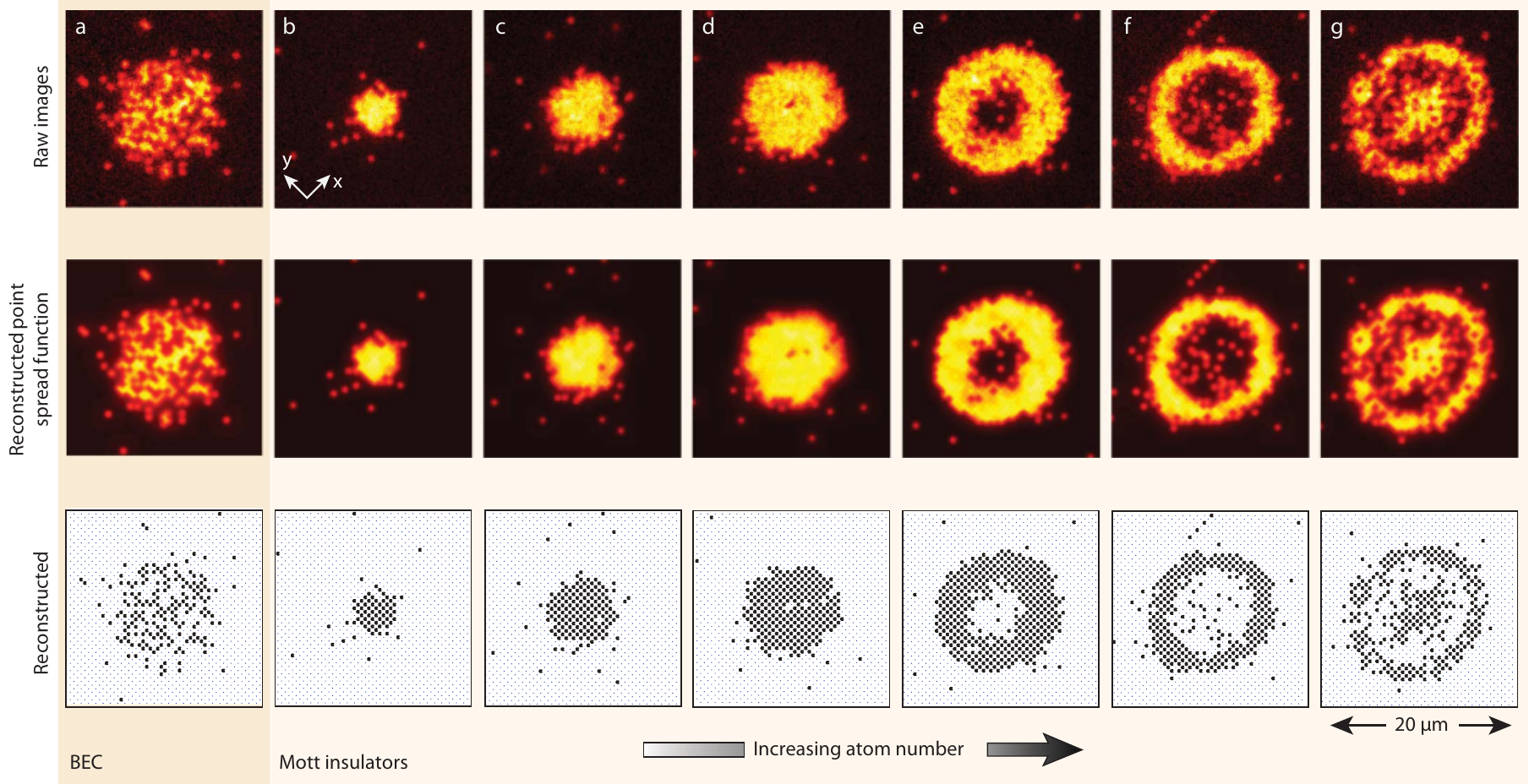}
\end{center}
\caption{High-resolution fluorescence images of a Bose-Einstein condensate and Mott
insulators in a two-dimensional optical lattice. The top row shows experimentally obtained raw images of a Bose-Einstein condensate
(a) and Mott insulators of rubidium atoms for increasing particle numbers (b-g). The middle row shows numerically reconstructed atom
distribution on the lattice. The images were convoluted with the point spread function. The bottom row shows the reconstructed atom number distribution. From Ref.~\cite{Sherson_2010}.}  \label{fig:Bloch_MI}
\end{figure}

\begin{figure}[t!]
\begin{center}
\includegraphics[width=0.6\textwidth]{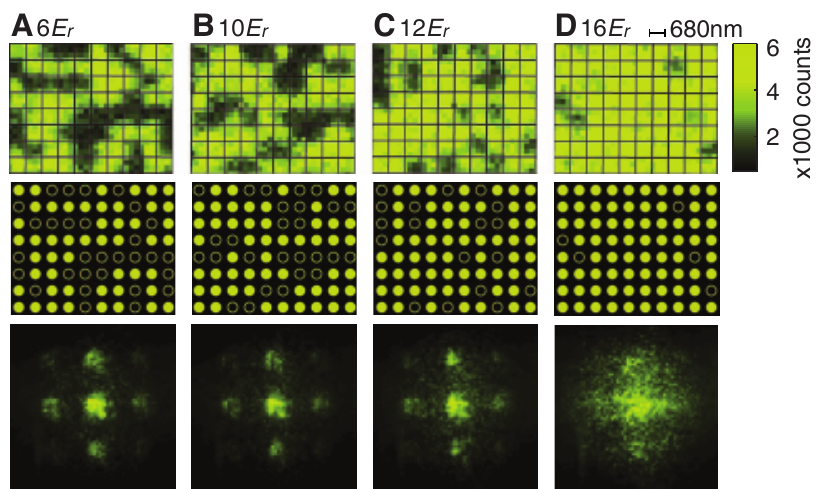}
\end{center}
\caption{Single-site imaging of atom number fluctuations across the superfluid-Mott insulator transition. (A to D) Top row: In situ fluorescence images of rubidium atoms from a region of $10\times 8$ lattice sites within the $n=1$ Mott shell that forms in a deep optical lattice. In the superfluid regime (A and B), sites can be occupied with odd or even atom numbers, which appear as full or empty sites, respectively, in the images. In the Mott insulator, occupancies other than 1 are highly suppressed (D). Middle row: reconstructed site occupancy. Solid and open circles indicate the presence and absence, respectively, of an atom on a site. Bottom row: Time-of-flight fluorescence images after 8-ms expansion. From Ref.~\cite{Bakr2010}.}  \label{fig:Greiner_MI}
\end{figure}

Single atom resolved imaging of a quantum gas in a two-dimensional optical lattice with sub-micrometer lattice spacing has been first demonstrated in Refs.\,\cite{Bakr_2009,Bakr2010,Sherson_2010} (see Fig.\,\ref{fig:Bloch_MI} and Fig.\,\ref{fig:Greiner_MI}). The imaging relies on the detection of fluorescence light, which is emitted by the atoms upon exposure the near-resonant light fields. Several constraints have to be taken into account:

\begin{itemize}

\item{The high resolution which is required to resolve submicrometer length scales requires an optical imaging system working at the diffraction limit.}

\item{The resulting short depth of focus requires the confinement of the atoms to a single plane.}

\item{Within that plane, the atoms have to be pinpoint to their position while they are imaged. This entails the use of a two-dimensional optical lattice.}

\item{As the atoms undergo many subsequent absorption and emission cycles, the applied light field are usually designed to provide laser cooling. The lattice potential has then to be deep enough to suppress thermal hopping during the imaging.}

\end{itemize}

\vspace{3cm}

\onecolumngrid

\twocolumngrid  

In these first experiments, which were all carried out with rubidium atoms, an optical molasses was used to provide the necessary cooling. During the optical molasses the atoms isotropically emit photons with scattering rate $\Gamma_\mathrm{scat}$ (see Eq.\,\ref{eq:scatteringrate}). The magnitude of $\Gamma_\mathrm{scat}$ depends on the parameters of the optical molasses. Typical values are on the order of $\Gamma_\mathrm{scat}=10^5$\,s$^{-1}$. A quantitative discussion of the fluorescence imaging of dense clouds of neutral atoms can be found in Refs.\,\cite{Cirac1996,Shotter2011}. The solid angle that is captured by the imaging system determines the number of collected photons per atom

\begin{equation}
N_{\mathrm{ph}}=\Gamma_\mathrm{scat}\Omega T \tau \eta, 
\end{equation}

where $\Omega=(\mathrm{NA})^2/4$ is the solid angle that is captured by the imaging system, $\mathrm{NA}$ is the numerical aperture, $T$ is the transmission of the optics, $\tau$ is the exposure time and $\eta$ is the quantum efficiency of the CCD camera which collects the photons. For a numerical aperture of $\mathrm{NA}=0.7$, which requires a dedicated imaging system close to the atoms, a few thousand photons per atom are detected within a typical exposure time of a few hundred milliseconds. This large number of photons translates into a detection fidelity of more than 99.5\,\%. The exposure time is ultimately limited by the appearance of background gas collisions.

During the imaging, the optical lattice potential has to be much deeper than the temperature set by the optical molasses. Typical temperatures are on the order of 10 - 20\,$\mu$K and the lattice potential has to be a factor of 10 stronger in order to keep the atoms confined. Expressed in terms of the recoil energy, $V_{\mathrm{lattice}}\approx 1000\times E_{\mathrm{recoil}}$. Such deep optical potentials require high power lasers (e.g. 100\,W YAG laser), focused down to a beam waist of a few tens of micrometers. If the optical lattice was too shallow, the atoms start hopping during the imaging procedure \cite{Bakr_2009}.

The optical molasses cooling discussed above does not work for all atomic species and different cooling techniques have to be used. In the recent experiments with single fermionic atoms, two different strategies have been pushed forward, both having their origin in laser cooling of trapped ions. Raman sideband cooling (Fig.\,\ref{fig:Raman}) has been applied to fermionic lithium atoms \cite{Parsons2015,Omran2015} and potassium atoms \cite{Cheuk2015}. Alternatively, electromagnetically induced transparency (EIT) cooling \cite{Morigi2000,Roos2000} has been used for potassium \cite{Haller2015,Edge2015}. In this scheme, the EIT condition in a $\lambda$-type of system prevents the absorption of a cooling laser, except for the situation where one vibrational quantum can be removed from the atom. In all three schemes, the scattered photons of the cooling scheme form again the basis for the imaging. For the imaging of single ytterbium atoms \cite{Miranda2015} no dedicated laser cooling scheme was used, but the imaging was made fast enough to prevent the atoms from hopping between the lattice sites.

Resolving the fluorescence of single atoms residing in a lattice with 500\,nm spacing is a major challenge. However, the requirements on the spatial resolution are a bit relaxed, if one applies the concept of the point spread function (PSF). It describes the intensity distribution of a point-like source in the imaging plane. The signal of an atom is then given by the convolution of the PSF with the atomic density distribution. As every atom is identical, the collected fluorescence of many atoms in the lattice is the sum over individual, identical signals. The spatial distribution of the atoms can then be reconstructed with high fidelity in a post processing step. This principle was first demonstrated in a one-dimensional optical lattice with 433\,nm period and sparse filling, where neighboring atoms could be identified despite a $\sigma$-width of the point-spread function of 810\,nm \cite{Karski2009}. The extension of this technique to two-dimensional lattices is now routinely applied.

As the depth of focus is limited to a few micrometers, essentially only two-dimensional samples are studied in the experiment. The atoms are either initially loaded in only one plane of a perpendicular optical lattice \cite{Bakr2010} or the experiment starts from a three-dimensional lattice configuration, from which all atoms but those in a single plane are removed \cite{Sherson_2010}. This is realized by applying a magnetic field gradient and performing a microwave frequency sweep, which flips the spins of the atoms in all planes. The plane of interest is then flipped back again by a resonant microwave pulse. The remaining atoms are subsequently removed by a resonant laser pulse. 

The resolution requirements can be relaxed if the lattice spacing is larger. In this case, even the full three-dimensional reconstruction of the atom distribution is possible by translating the objective with a piezoelectric actuator and imaging all planes successively \cite{Nelson_2007}. The detection of atoms in an optical lattice for a lattice spacing of 2\,$\mu$m, where tunneling is present, has been reported in Ref.\,\cite{Itah2010}. As tunneling for lattice constants larger than one micrometer is rather slow, such approaches are more relevant in the context of quantum information processing.

\begin{figure}[t!]
\begin{center}
\includegraphics[width=0.395\textwidth]{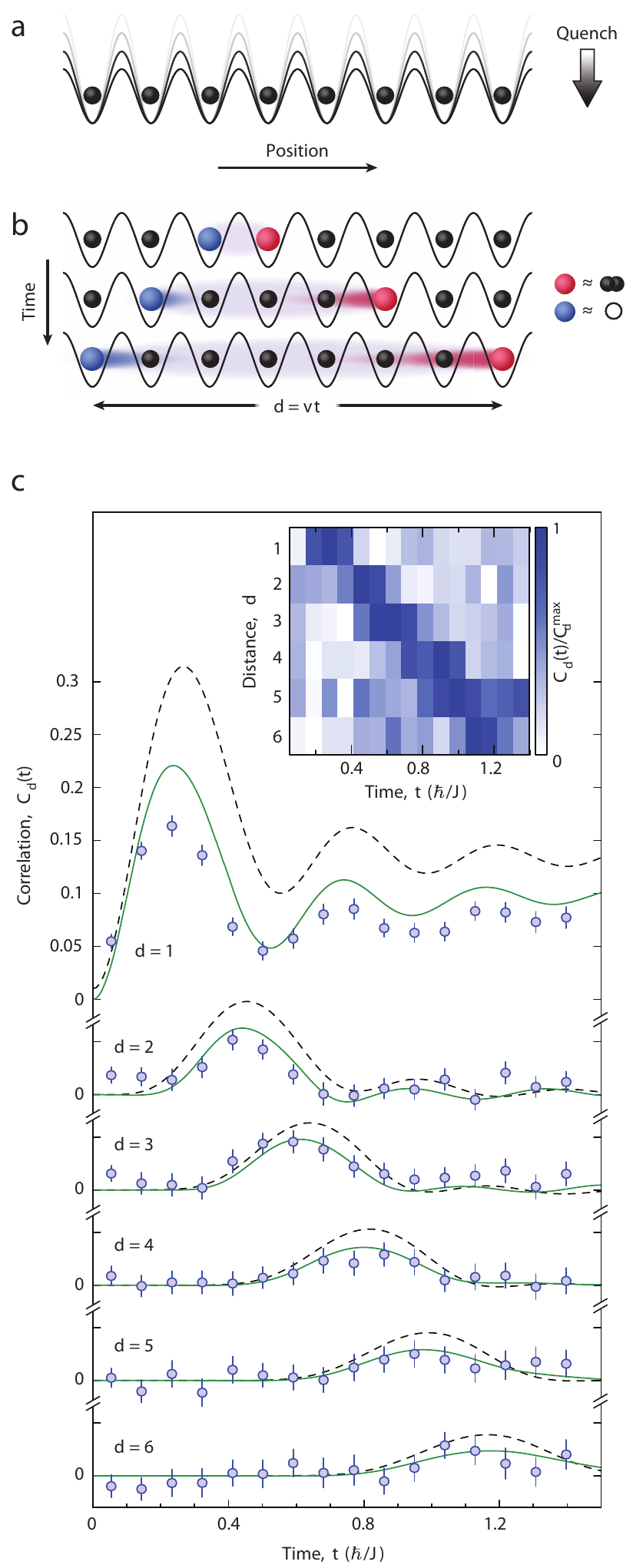}
\end{center}
\caption{Spreading of correlations in a quenched atomic Mott insulator. (a) A one-dimensional atomic Mott insulator is quenched to a lower lattice depth. (b) Entangled doublon-holon pairs emerge at all sites and propagate ballistically in opposite directions. 
(c) A positive correlation signal propagates with increasing time to larger distances. The integer $d$ denotes the spatial separation of the lattice sites for which the correlations are evaluated. The experiment is in good agreement with numerical simulation for an infinite´, homogeneous system at zero temperature (continuous green line). The dashed line is a simplified analytical model. From Ref.~\cite{Cheneau2012}.}  \label{fig:lightcone}
\end{figure} 

An important characteristic of fluorescence imaging in an optical lattice is the so-called parity-projection. As two atoms at the same lattice site can rapidly undergo a light-assisted collision \cite{Sompet2013}, they disappear at the very beginning of the imaging sequence. Lattice sites with more than two atoms have either one atom left (odd initial atom number) or no atom left (even initial atom number). The detection can also be made internal state-sensitive through the combination of a controlled push out of atoms in one internal state with the detection of the remaining atoms in the other internal state.

The successful realization of single atom fluorescence imaging in a quantum gas, has allowed for the first study of single atom resolved bosonic Mott insulators \cite{Sherson_2010,Bakr2010}. Fig.\,\ref{fig:Bloch_MI} and Fig.\,\ref{fig:Greiner_MI} reveal the high level of single atom control in these experiments. Shortly afterwards, a series of ground-breaking experiments of many-body quantum phases and correlated many-body quantum dynamics have been performed. In the following we briefly review two experiments in order to highlight the potential of this technique.

In the first experiment to discuss, the propagation speed of particle correlations in a strongly interacting quantum gas has been investigated \cite{Cheneau2012}, see Fig.\,\ref{fig:lightcone}. It has been predicted that in such systems, a maximal velocity for the propagation of correlations, known as the Lieb-Robinson bound, exists \cite{Lieb1972}. As a consequence, correlations spread lightcone-like. The experiment was preformed as follows: starting from a one-dimensional bosonic Mott insulator, a quantum quench to a lower lattice height has been used to create doublon-holon pairs, which propagate through the one-dimensional chain. The holon and doublons are correlated with each other and propagate in opposite directions. The correlation length between them thus increases in time. Fig.\,\ref{fig:lightcone} shows how the correlations propagate in space, evidencing the existence of a maximum velocity. The experimentally determined velocity agrees well with numerical simulations.

The second experiment is an example for the quantum simulation of an antiferromagnetic spin-1/2 chain \cite{Simon2011} with the help of strongly interacting bosons in a tilted optical lattice. The model Hamiltonian for this system is given by

\begin{equation}\label{eq:AIM}
H_\mathrm{Ising}=J\sum_i \left( S_z^i S_z^{i+1} - h_z^i S_z^i - h_x^iS_x^i\right).
\end{equation} 

\begin{center}
\begin{figure}[h!!!]
\includegraphics[width=0.75\textwidth]{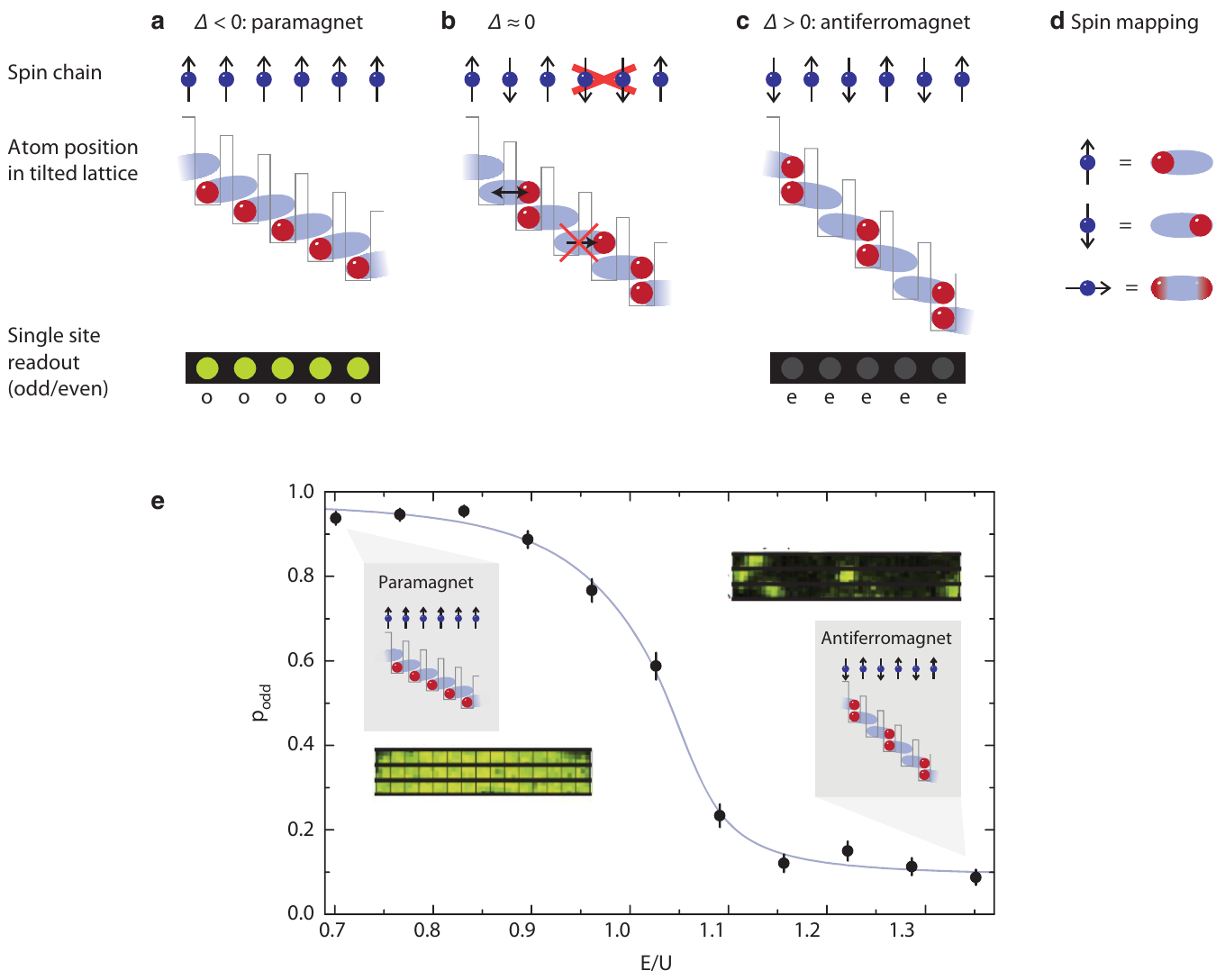}

\caption{Mapping of the Bose-Hubbard model onto the Ising spin chain. (a) A tilted Mott insulator maintains unity occupancy until the offset energy between two neighboring sites equals the on-site interaction energy U ($\Delta \leq 0$). (b) For $\Delta=0$ an atom can tunnel to the neighboring site if the atom on that site has not itself tunneled. (c) Tilting further ($\Delta>0$), the system undergoes a transition into a doubly degenerate staggered phase. (d) The system maps onto interacting spin-1/2 particles,
whose two spin states correspond to the two possible locations of each atom. The tunnelling constraint forbids adjacent down spins, realizing an effective spin-spin interaction. Top row: effective spin pattern. Bottom row: odd/even signature in the fluorescence image. (e) Phase transition from a paramagnet to an antiferromagnet, driven by the external magnetic field (realized as tilt of the optical lattice). The two insets show the corresponding fluorescence images for the paramagnet (left, every site is singly occupied) and anti-ferromagnet (right, no atom or two atoms per site, resulting in the absence of fluorescence). From Ref.~\cite{Simon2011}.}  \label{fig:AFSC}
\end{figure}
\end{center}

Here, $S_x^i$ and $S_z^i$ denote the spin projection operators along the $x$- and $z$-direction at site $i$, $J$ is the coupling strength between nearest neighbor spins and $h_z^i$ ($h_x^{i}$) are the $z$($x$)-component of the external magnetic field at site $i$. In the presence of a constant force, the Bose-Hubbard model (\ref{eq:BHM}) can be mapped  on the spin Hamiltonian (\ref{eq:AIM}) following the recipe described in Ref.\,\cite{Sachdev2002}. Fig.\,\ref{fig:AFSC} illustrates the microscopic physics on the level of bosonic atoms in a tilted optical lattice and how effective spins arise from the interaction blockade between the atoms. In order to drive the system between the different quantum phases, the tilt, which corresponds to the externally applied magnetic field, is changed. As a consequence, the system changes from a paramagnetically ordered state (one atom at each site) to an antiferromagnetic state (two atoms at every other site). Both phases can be discriminated with the help of the parity projection technique.

More experiments on many-body quantum systems with single atom fluorescence detection techniques include the observation of string order \cite{Endres2011} the realization of algorithmic cooling \cite{Bakr2011}, photo-assisted tunneling in a strongly correlated Bose gas \cite{Ma2011}, the study of a 'Higgs' amplitude mode \cite{Endres2012}, the observation of magnon bound states \cite{Fukuhara2013a}, the dynamics of a spin impurity \cite{Fukuhara2013}, spin transport in Heisenberg quantum magnets \cite{Hild2014}, and the measurement of entanglement entropy \cite{Islam2015}. New developments include the simultaneous imaging of two spin states \cite{Preiss2015b,Fukuhara2015}, and the imaging of more than one plane of atoms in combination with spin-resolved readout \cite{Preiss2015b}.

\onecolumngrid

\twocolumngrid

\subsubsection{Dispersive coupling to a cavity}

The interaction of an atom with a light field can be enhanced if the atom is located inside an optical cavity. The interaction with the light field is then described by the Jaynes-Cummings Hamiltonian \cite{Jaynes1963}

\begin{equation}
H_\mathrm{JC}=\hbar \omega_0 \sigma^+ \sigma^- + \hbar \omega_L a^\dagger a + \hbar g\left( a^\dagger \sigma^- + a \sigma^+ \right),
\end{equation}

where $\omega_0$ is the transition frequency of the atom, $\omega_L$ is the cavity frequency, $\sigma^+$ ($\sigma^-$) are the raising (lowering) operators of the atom, $a^\dagger$ ($a$) denotes the creation (annihilation) operator of a cavity photon and $g$ is the atom-photon coupling strength. The atom-light coupling leads to the formation of dressed states, which are shifted in energy. The presence of an atom in the cavity therefore shifts the resonance frequency of the cavity and the transmission of a weak probe beam is changed. This has been used to detect single atoms outcoupled from a Bose-Einstein condensate. The subsequent arrival times of the atoms has allowed, e.g., to measure the atom number statistics of an atom laser \cite{Öttl2005}. 

When a  Bose-Einstein condensate is located inside the cavity, long-range interactions between the atoms can be mediated by the photons in a cavity mode. As the resonator enables the photons to strongly interact with all atoms, long-range order can build up. In this context, the Dicke Hamiltonian \cite{Dicke1954,Tavis1968} has been studied \cite{Baumann2010}. It describes the collective coupling of $N$ two-level systems, which form a collective spin $\vec{J}=\sum\vec{j}_i$, to a single light mode. For increasing drive with a laser perpendicular to the cavity, the system undergoes a phase transition from a superfluid phase to a self-organized phase. This phase is characterized by a checkerboard pattern, where every second site is occupied by an atom. While this technique has no access to individual atoms of the system, the emerging photons from the cavity can be used to probe the system. In this way, atom number fluctuations become observable in real time \cite{Brennecke2013}. The dynamical coupling of atoms to cavities has been reviewed in Ref.\,\cite{Ritsch2013}. Ultracold atoms coupled to a cavity have also been used to demonstrate many-body entanglement \cite{McConnell2015,Haas_2014}.

\subsubsection{Absorption imaging}

While optical single atom detection is mainly based on fluorescence imaging, the detection of large numbers of atoms is usually done in time of fight absorption imaging. Absorption imaging on the single atom basis is not a standard technique; however, it should be noted that it is possible to see the absorption signal of a single atom \cite{Tey2008}. In the context of quantum gas research, single atom sensitivity has not been reached so far. However, a careful noise analysis of absorption images unravels the underlying atom fluctuations. This has led to the observation of correlations in a bosonic Mott insulator \cite{Foelling_2005} and anticorrelations \cite{Rom2007} in a fermionic Mott insulator state. While these experiments were performed in time of fight, \textit{in situ} detection of atom number fluctuations has been demonstrated for 2D atomic samples. This has been used to measure density fluctuations in a Mott insulator \cite{Gemelke2009} and to observe scale-invariance in a two-dimensional Bose gas \cite{Hung2011}.

\subsection{Generalized measurements}
The above presented detection techniques can be understood as a projective measurement of the atom number distribution. The interpretation of the measurement result is therefore rather clear. However, the possibility to continuously observe the system under the influence of the measurement of a part of the system requires a more careful description of the detection process. In this section, we look in the details of the measurement process in fluorescence imaging and in scanning electron microscopy. For the sake of simplicity, we consider a lattice model. An ideal detection process measures the exact number of atoms in a given lattice site. It can therefore be described with von Neumann projectors:

\begin{equation}
\hat{P}_{n_i}=\ket{n_i}_{i\,i}\hspace{-0.08cm}\bra{n_i},
\end{equation}

where $n_i$ denotes the occupation number in lattice site $i$. We now assume that all lattice sites are measured simultaneously. The projection onto a particular realization of occupation numbers $\ket{\{ n_i \} }=\ket{n_1,...,n_k}$, is then described by the operator

\begin{equation}
\hat{P}_{\{n_i\}}=\prod_i^k \ket{n_i}_{i\,i}\hspace{-0.08cm}\bra{n_i}=\ket{n_1}_{1\,1}\hspace{-0.08cm}\bra{n_1}\otimes ... \otimes \ket{n_k}_{k\,k}\hspace{-0.08cm}\bra{n_k}.
\end{equation}

The density operator $\rho$ of the many-body system is involved into the mixed state

\begin{equation}\label{eq:rho}
\rho'=\sum_{\{n_i \}}\hat{P}_{\{n_i\}} \rho \hat{P}_{\{n_i\}},
\end{equation}

and the density matrix for a particular outcome of the measurement with atom distribution $\{ n_i \}$ is given by

\begin{equation}\label{eq:rho2}
\rho_{\{n_i \}}=\frac{\hat{P}_{\{n_i\}} \rho \hat{P}_{\{n_i\}}}{Tr[\rho \hat{P}_{\{n_i\}}]   }.
\end{equation}

The fluorescence imaging of a many-body system in a lattice differs from this description as parity projection limits the outcome of the measurement in each site to the sub-space $\{\ket{0},\ket{1}\}$. This difference can be captured by an extended description in terms of generalized measurement operators \cite{Wiseman_2010} for each site:

\begin{eqnarray}
\hat{M}_{\mathrm{n_i}}=\ket{0}_{i\,i}\hspace{-0.08cm}\bra{n_i}, n_i=\mathrm{even}\\
\hat{M}_{\mathrm{n_i}}=\ket{1}_{i\,i}\hspace{-0.08cm}\bra{n_i}, n_i=\mathrm{odd},
\end{eqnarray}

The density matrix is then evolved under the measurement in analogy to equations (\ref{eq:rho}) and (\ref{eq:rho2}):

\begin{equation}
\rho'=\sum_{\{n_i \}}\hat{M}_{\{n_i\}} \rho \hat{M}^{\dagger}_{\{n_i\}}
\end{equation}

and

\begin{equation}
\rho_{\{n_i \}}=\frac{\hat{M}_{\{n_i\}} \rho \hat{M}^{\dagger}_{\{n_i\}}}{Tr[ \rho \hat{M}_{\{n_i\}} \hat{M}^{\dagger}_{\{n_i\}}]   }.
\end{equation}

with

\begin{equation}
\hat{M}_{\{n_i\}}=\prod_i^k \hat{M}_{n_i}.
\end{equation}

The above discussion is rather academic if the system is not further evolved in time after the measurement. However, scanning electron microscopy allows for a local measurement of the occupation number by means of partial atom removal while the remaining system evolves in time. The action of removing a particle by electron impact ionization (or by any other loss process) corresponds to the local application of the annihilation operator $\hat{a}$. For a generalization to the detection in continuous space, see for instance Ref.\,\cite{Barontini_2013}. The rate at which the atoms are removed from the lattice site depends on the chosen parameters and is denoted by $\gamma$. We assume that the site is illuminated for a short time $\Delta t$. Two physical outcomes after the time $\Delta t$ are possible: the detection of a particle as evidenced by a click in the detector or the non-detection of a particle, signaled by the absence of a detector event. The two corresponding operators are defined as \cite{Wiseman_2010}

\begin{eqnarray}
\hat{M}_
1&=&\sqrt{\epsilon}\hat{a},\,\,\,\,\,\,\,\,\,\,\,\,\,\,\textrm{detection of a particle}\\
\hat{M}_2&=&1-\frac{\epsilon}{2}\hat{a}^\dagger\hat{a}\,\,\,\,\textrm{non-detection of a particle}
\end{eqnarray}

with 

\begin{equation}
\hat{M}_1^\dagger \hat{M}_1 + \hat{M}_2^\dagger \hat{M}_2 = \mathbb{1},
\end{equation}

provided that the probability for the detection of a particle, given by $\epsilon=\gamma\Delta t$, is much smaller than 1. The density matrix is evolved according to

\begin{equation}
\rho'=\sum_i \hat{M}_i \rho \hat{M}_i^\dagger=\hat{M}_1 \rho \hat{M}_1^\dagger+\hat{M}_2 \rho \hat{M}_2^\dagger
\end{equation}

The differential change in the density matrix after the time $\Delta t$ is given by

\begin{eqnarray}
\Delta \rho &=& \rho' - \rho \\
&=& \epsilon \hat{a} \rho \hat{a}^\dagger + \left[ 1-\frac{\epsilon}{2}\hat{a}^\dagger\hat{a}\right] \rho \left[1-\frac{\epsilon}{2}\hat{a}^\dagger\hat{a}\right] -\rho\\
&=&  \gamma \Delta t \hat{a} \rho \hat{a}^\dagger - \frac{1}{2}\gamma\Delta t \hat{a}^\dagger\hat{a} \rho - \frac{1}{2}\gamma\Delta t \rho \hat{a}^\dagger\hat{a}.
\end{eqnarray}

In the limit $\Delta t \rightarrow 0$, the last equation converts into a differential equation for the density matrix and takes the well known form of a master equation in Lindblad form

\begin{eqnarray}
\dot{\rho}&=&\mathcal{L}\rho,\hspace{0.5cm} \textrm{with} \\ \mathcal{L}\rho &=& -\frac{i}{\hbar}\left[H,\rho\right]+\frac{\gamma}{2}\left( 2 \hat{a} \rho \hat{a}^\dagger - \hat{a}^\dagger\hat{a} \rho - \rho\hat{a}^\dagger\hat{a} \right),
\end{eqnarray}

where we have additionally introduced a unitary time evolution under the Hamiltonian $H$. Locally detecting atoms is therefore a way to investigate open many-body quantum systems. For strong enough detection strength, the non-unitary part of the master equation can dominate the unitary time evolution. The back action of the measurement process then becomes important and one enters the regime of the quantum Zeno effect \cite{Itano1990,Fischer2001}, where the coherent time evolution is modified by the measurement. In ultracold quantum gases, theoretical studies \cite{Brazhnyi_2009,Barmettler2011,Witthaut2011} and first experimental work on dissipative \cite{Barontini_2013} and driven-dissipative \cite{Labouvie2015b} Bose-Einstein condensates as well as the observation of the quantum Zeno effect in a three-dimensional optical lattice \cite{Patil2015} show how single particle detection is connected to the control of open quantum systems. In the weak probing limit, where the back action of the measurement process can be neglected ($\hbar\gamma\ll||H||$), the subsequent detection of the particles can be used to measure time-dependent correlation functions locally. This has been applied to temporal pair correlations in thermal gases \cite{Guarrera_2011} and one-dimensional quantum gases \cite{Guarrera_2012}.

\section{Microscopic manipulation of atoms in a quantum gas}
The spatially resolved manipulation of atoms in a quantum gas is in many cases directly connected to the ability to detect atoms locally. The manipulation can be implemented as density engineering, where atoms from specific lattice sites or regions are removed from the system, or by coherently changing the internal state of one or more atoms at given locations. High resolution \textit{in situ} density engineering has been realized by local removal of atoms by electron impact \cite{Wurtz_PRL_2009,Labouvie2015} or by applying local spin flip operations followed by the removal of one spin state with an optical push beam  \cite{Weitenberg2011,Fukuhara2013,Preiss2015}. Density engineering is a strategy to induce a non-equilibrium initial condition in a quantum system. The ensuing dynamics after such a quench allows for the study of the tunneling dynamics of single atoms \cite{Weitenberg2011} and strongly correlated quantum walks \cite{Preiss2015} as well as many-body transport phenomena such as negative differential conductivity \cite{Labouvie2015} or driven dissipative superfluids \cite{Labouvie2015b}. In optical lattices, the density quench is typically realized in a frozen lattice, where tunneling is absent. The dynamics is then induced by lowering the lattice depth to a value where tunneling sets in (see also II.B.1).

\begin{figure}[b!]
\begin{center}
\includegraphics[width=0.45\textwidth]{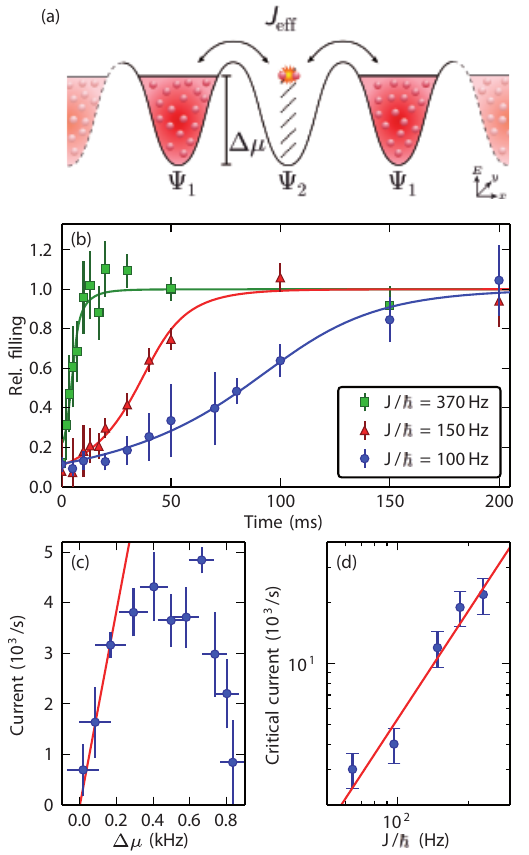}
\end{center}
\caption{Non-equilibrium dynamics of a Bose-Einstein condensate in a one-dimensional optical lattice with 600\,nm lattice spacing. The initial density distribution is that of Fig.\,\ref{fig:ebeam}c, top row. A full site contains about 700 atoms. (a) Microscopic level structure: the chemical potential $\mu$ of a full site is larger than the radial vibrational energy. (b) Temporal evolution of the site occupancy for different tunneling couplings $J$. (c) The current-voltage characteristics, which is given by the derivative of the data shown in (b), shows negative differential conductivity (NDC). (d) The dependence of the maximum current on the tunneling coupling scales quadratically in $J$, indicating an incoherent hopping transport. Taken from Ref.~\cite{Labouvie2015}.}  \label{fig:NDC}
\end{figure}

The following example shows how high resolution density engineering can be used to study non-equilibrium dynamics and mass transport in many-body quantum systems. Fig.\,\ref{fig:NDC} shows how a weakly interacting Bose-Einstein condensate refills an empty lattice site in a one-dimensional optical lattice with high filling (700 atoms per lattice site). Because the atoms occupy many radial modes ($\omega_r=2\pi\times170$\,Hz, $\mu=2\pi\times1500$\,Hz) the refilling dynamics is non-trivial. Due to a combination of intrinsic collisions and non-linear tunneling coupling the resulting current-voltage characteristics exhibit negative differential conductivity \cite{Labouvie2015}.

\begin{figure}[t!]
\begin{center}
\includegraphics[width=0.5\textwidth]{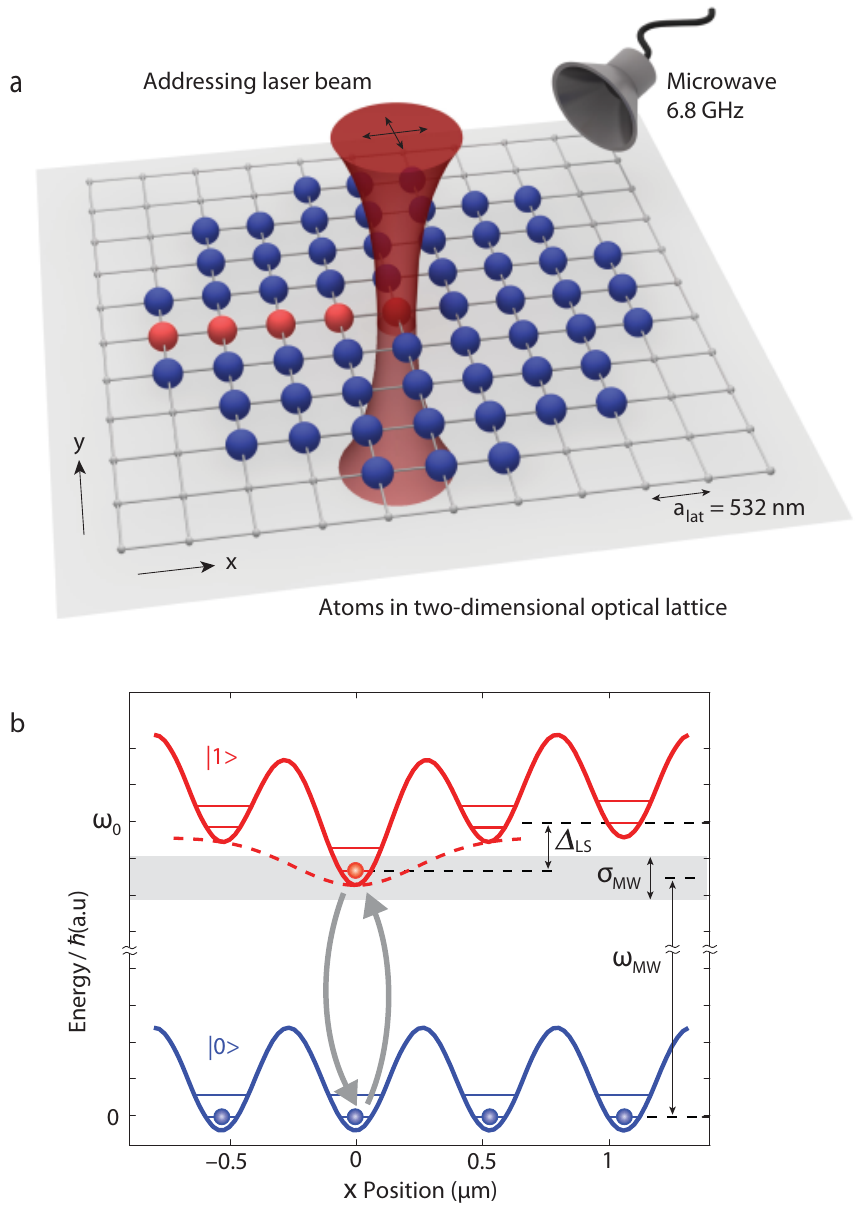}
\end{center}
\caption{Coherent manipulation of single atoms in an optical lattice. (a) A focussed laser beam creates a differential light shift for two different hyperfine states, which act as an effective two-level system. (b) A microwave pulse induces the spin flip. Taken from Ref.\,\cite{Weitenberg2011}} \label{fig:addressing}
\end{figure}
 
Spin flip operations of single atoms have been realized with localized light shifts in experiments employing fluorescence imaging \cite{Weitenberg2011,Fukuhara2013}. The already existing optical microscope is used to overlap an additional addressing beam which is focused onto the atomic sample by the imaging system (Fig.\,\ref{fig:addressing}). With the help of piezo mirrors, the beam is scanned across the atomic cloud. The coherent operations are performed as follows: in a first conceptual step, a pseudo-spin 1/2 system is encoded in two different hyperfine states. A convenient choice for rubidium is $\ket{\uparrow}=\ket{F=1,m_F=-1}$ and $\ket{\downarrow}=\ket{F=2,m_F=-2}$. Transitions between the two hyperfine states are driven by microwave radiation at a frequency of 6.8\,GHz. In order to select a specific lattice site, a focused laser is pointed at the particular site (Fig.\,\ref{fig:addressing}a). The polarization of the laser beam is chosen such that the two hyperfine states experience different light shifts and the transition frequency between the two states is shifted locally. Spin flips are then driven by frequency sweeps of the microwave radiation \cite{Weitenberg2011} (Landau Zener transition) or by the application of a resonant microwave pulse \cite{Fukuhara2013}. In combination with a push-out laser beam, spin flip operations are used to remove atoms from specific sites in an optical lattice as described above. Local spin flips have been used to study magnon bound states \cite{Fukuhara2013a} and the dynamics of spin impurities \cite{Fukuhara2013}.

An alternative approach can be realized with a spatial light modulator such as a digital mirror device (DMD). A DMD is a two-dimensional array of micrometer-sized mirrors ($10-15\,\mu$m size), which can be individually flipped. Directly imaging a laser beam, which is incident on the DMD, onto the atoms \cite{Fukuhara2013} or manipulating the intensity in the Fourier plane \cite{Preiss2015} allows for the site-resolved application of additional light fields at each lattice site. To set the intensity in the lattice sites smoothly, the DMD is demagnified on to the atoms such that the signal from many mirrors contributes to the light field at a single site. This approach has the advantage of addressing several sites at the same time and can also be used to generate arbitrary potential landscapes, which can be used, e.g., for building atomtronic circuits \cite{Pepino2009,Eckel2014} or applying disorder.

The combination of single atom detection and local coherent manipulation allows in principle for a full quantum state tomography. Experiments with ions have a long history in this context and have recently demonstrated the tomography of the quantum dynamics of an 8 qubit spin system \cite{Jurcevic2014}. Ultracold atoms have not yet employed such large scale tomography sequences, but the successful addressing schemes show that this a viable route.

\section{Future perspectives}

The successful detection of single atoms in a quantum gas has stimulated the research field of ultracold atoms in the past years. Most notably, fluorescence imaging has now been extended to fermionic atoms \cite{Cheuk2015,Haller2015,Parsons2015,Omran2015,Edge2015}. In many laboratories, related approaches to image bosonic and fermionic atoms with high sensitivity and high spatial resolution are under way and the next years will likely see an explosion of experimental activities in this direction. The high level of control that is now available in the experiment opens the door to an in-depth study of many-body quantum systems. This includes, e.g., the exploration and characterization of new quantum phases in optical lattices, the non-equilibrium dynamics of many-body quantum systems, the investigation of open quantum systems, lattice physics of bosons and fermions in higher orbitals, the exploration of novel lattice geometries, the creation and characterization of topological edge states and interacting many-body quantum walks as well as the creation of microscopic atomtronic circuits. We close this report by briefly highlighting three examples where further advancing the existing technology pushes state-of-the-art quantum research. 

\subsection{Fermionic lattice gases}

\begin{figure}[t!]
\begin{center}
\includegraphics[width=0.45\textwidth]{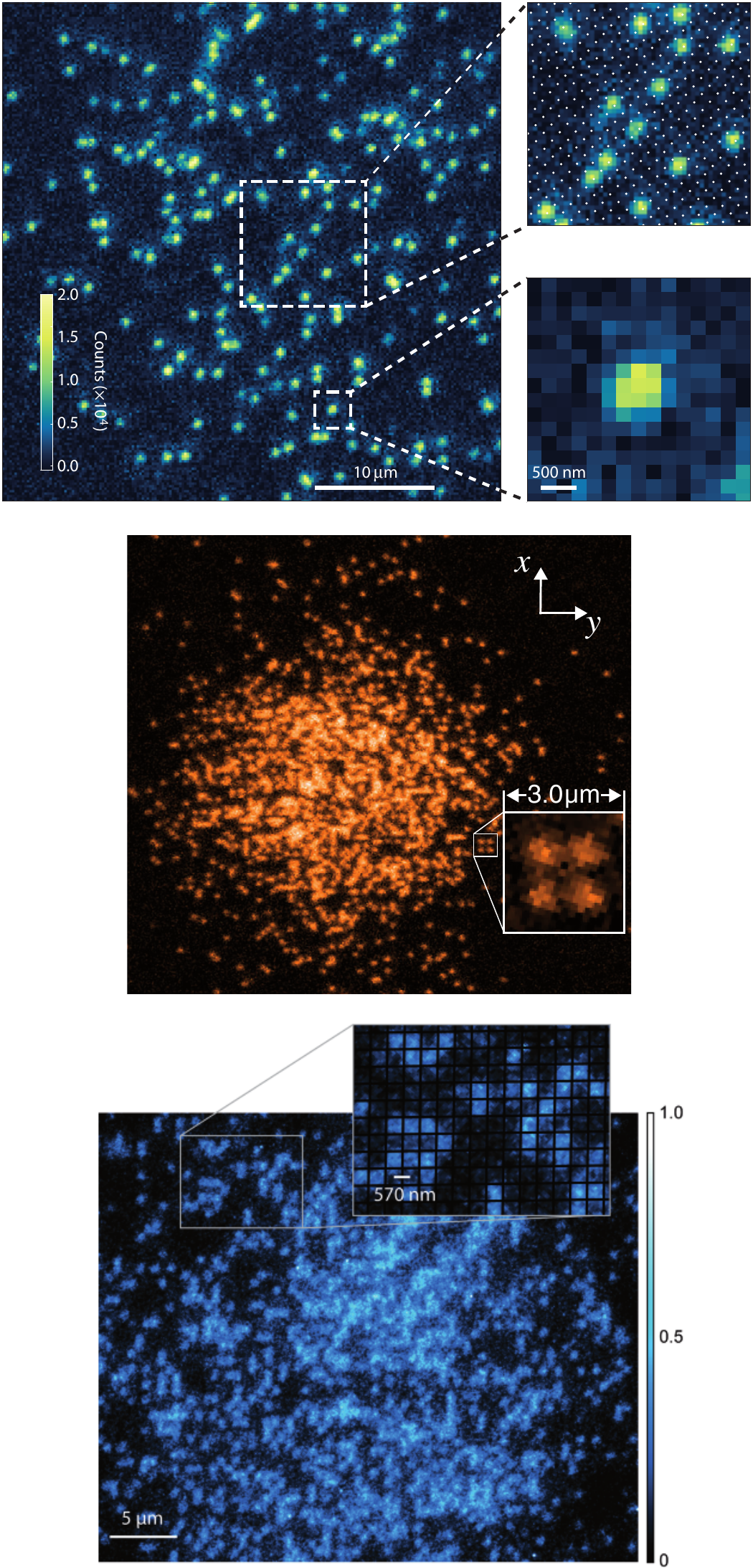}
\end{center}
\caption{Detection of fermionic atoms in a two-dimensional optical lattice. Top: $^{40}$K atoms in a lattice with 532\,nm spacing \cite{Haller2015}. Middle: $^{40}$K atoms in an optical lattice with 541\,nm spacing \cite{Cheuk2015}. Bottom: $^6$Li atoms in an optical lattice with 570\,nm spacing \cite{Parsons2015}.}  \label{fig:fermions}
\end{figure}

Directly after the successful imaging of bosonic atoms in quantum gases, the quest for single atom detection in fermionic quantum gases started. Five groups have now successfully imaged single fermionic $^{40}$K  \cite{Haller2015,Cheuk2015,Edge2015} and $^6$Li atoms \cite{Parsons2015,Omran2015} in a two-dimensional optical lattice with single site resolution (see Fig.\,\ref{fig:fermions}). An in depth understanding of fermionic many-body quantum systems is essential to predict and design the properties of strongly correlated materials. Their complex physics is also responsible for some of the most urgent unsolved problems in condensed matter physics \cite{Dagotto2005}, most notably the lack of a proper understanding of high-Tc superconductivity. One main reason for these problems is the antisymmetric wave function of fermionic many-body systems, which poses severe restrictions on the feasibility of numerical simulations (see, e.g., the so-called sign problem in quantum Monte Carlo simulations \cite{Loh1990}). Quantum gas microscopy of fermionic atoms will allow for a detailed study of many of those phenomena with direct access to correlation functions and their dynamics. Such experiments will help to benchmark theoretical models, ultimately unraveling the puzzling nature of strongly correlated electron systems.

\subsection{Designing local interactions}

Having high resolution optical access to an ultracold atomic quantum system can be used to create tailored optical potentials and Hamiltonians with spatially varying atom-light coupling. The implementation of fully structured two-dimensional optical potential landscapes with the help of spatial light modulators \cite{Brandt2010,Bowman2015,Preiss2015b} allows for the implementation of atomtronic circuits \cite{Pepino2009,Eckel2014}. The same technique can also be used to locally change the interaction between the atoms: the discovery of optical Feshbach resonances \cite{Fedichev1996,Bohn1997,Theis2004,Enomoto2008,Bauer2009,Blatt2011,Yan2013,Fu2013} allows for such a spatial variation of the interatomic scattering length \cite{Yamazaki2010}. Recently, it has been demonstrated for  cesium atoms that the scattering length around a Feshbach resonance can be locally tuned with a light field at a particular (also called ''magic'') wavelength \cite{Clark2015}. In this way, boundaries between different quantum phases can be investigated, providing a link to edge states, multilayer structures and surface science.

\subsection{One-way quantum computing}

Single atom detection and manipulation is also appealing in the context of quantum information processing. Two-dimensional optical lattices with unit filling have been proposed as a resource for what is known as one-way quantum computing \cite{Raussendorf_2001}. In this approach, a lattice system of single atoms is initially entangled \cite{Mandel2003}. The quantum computation task is then performed by sequential local single-particle measurements in different basis. Thereby, the need for entangling two-particle quantum gates is replaced by the initial global entangling operation. Local measurements of the spin state with high fidelity but vanishing influence on the remaining atoms has not yet been shown. However, more complex atoms with metastable internal states or dedicated experimental sequences might overcome this problem, thus opening yet another promising research direction.

\section{Summary and conclusions}
The research field of ultracold quantum gases has seen several breakthroughs in the last decades. Starting from weakly interacting Bose-Einstein condensates and their excitation spectrum, quantum phase transitions in optical lattices and strongly interacting fermionic quantum gases have been explored. Molecular condensates and Fermi gases are also available. The latest developments include long-range interactions and artificial gauge fields. Single atom detection has proven to provide unprecedented insight and understanding of such systems revealing their microscopic structure, dynamics and correlations. Given the experimental progress over the last years, we can look forward to many more exciting results to come.

\section*{Acknowledgements}

I gratefully acknowledge fruitful discussions with A. Widera, M. Cheneau and C. Westbrook. Part of this work was supported by the DFG SFB/TR49.

\bibliographystyle{natbib}
\bibliography{references-HO}

\begin{thebibliography}{}

\bibitem[Adams and Riis(1997)Adams and Riis]{Adams1997}
Adams, C.~S. and Riis, E. (1997).
\newblock Laser cooling and trapping of neutral atoms.
\newblock {\em Prog. Quant. Electr.}, {\bf 21}, 1.

\bibitem[Alt {\em et~al.}(2003)Alt, Schrader, Kuhr, M\"uller, Gomer, and
  Meschede]{Alt2003}
Alt, W., Schrader, D., Kuhr, S., M\"uller, M., Gomer, V., and Meschede, D.
  (2003).
\newblock Single atoms in a standing-wave dipole trap.
\newblock {\em Phys. Rev. A\/}, {\bf 67}, 033403.

\bibitem[An {\em et~al.}(1994)An, Childs, Dasari, and Feld]{An1994}
An, K., Childs, J.~J., Dasari, R.~R., and Feld, M.~S. (1994).
\newblock Microlaser: A laser with one atom in an optical resonator.
\newblock {\em Phys. Rev. Lett.}, {\bf 73}, 3375.

\bibitem[Anderlini {\em et~al.}(2004)Anderlini, Courtade, Ciampini, M\"uller,
  Morsch, and Arimondo]{Anderlini2004}
Anderlini, M., Courtade, E., Ciampini, D., M\"uller, J.~H., Morsch, O., and
  Arimondo, E. (2004).
\newblock Two-photon ionization of cold rubidium atoms.
\newblock {\em J. Opt. Soc. Am. B\/}, {\bf 21}, 480.

\bibitem[Anderson {\em et~al.}(1995)Anderson, Ensher, Matthews, Wieman, and
  Cornell]{Anderson_1995}
Anderson, M., Ensher, J., Matthews, M., Wieman, C., and Cornell, E. (1995).
\newblock Observation of {B}ose-{E}instein condensation in a dilute atomic
  vapor.
\newblock {\em Science\/}, {\bf 269}, 198.

\bibitem[Bakr {\em et~al.}(2009)Bakr, Gillen, Peng, F\"olling, and
  Greiner]{Bakr_2009}
Bakr, W.~S., Gillen, J.~I., Peng, A., F\"olling, S., and Greiner, M. (2009).
\newblock A quantum gas microscope for detecting single atoms in a
  {H}ubbard-regime optical lattice.
\newblock {\em Nature\/}, {\bf 462}, 74.

\bibitem[Bakr {\em et~al.}(2010)Bakr, Peng, Tai, Ma, Simon, Gillen, F\"olling,
  Pollet, and Greiner]{Bakr2010}
Bakr, W.~S., Peng, A., Tai, M.~E., Ma, R., Simon, J., Gillen, J., F\"olling,
  S., Pollet, L., and Greiner, M. (2010).
\newblock Probing the superfluid-to-mott insulator transition at the
  single-atom level.
\newblock {\em Science\/}, {\bf 329}, 547.

\bibitem[Bakr {\em et~al.}(2011)Bakr, Preiss, Tai, Ma, Simon, and
  Greiner]{Bakr2011}
Bakr, W.~S., Preiss, P.~M., Tai, M.~E., Ma, R., Simon, J., and Greiner, M.
  (2011).
\newblock Orbital excitation blockade and algorithmic cooling in quantum gases.
\newblock {\em Nature\/}, {\bf 480}, 500.

\bibitem[Barmettler and Kollath(2011)Barmettler and Kollath]{Barmettler2011}
Barmettler, P. and Kollath, C. (2011).
\newblock Dynamical response of a bosonic quantum gas to local one-body losses.
\newblock {\em Phys. Rev. A\/}, {\bf 84}, 041606.

\bibitem[Barontini {\em et~al.}(2013)Barontini, Labouvie, Stubenrauch, Vogler,
  Guarrera, and Ott]{Barontini_2013}
Barontini, G., Labouvie, R., Stubenrauch, F., Vogler, A., Guarrera, V., and
  Ott, H. (2013).
\newblock Controlling the dynamics of an open many-body quantum system with
  localized dissipation.
\newblock {\em Phys. Rev. Lett.}, {\bf 110}, 035302.

\bibitem[Barredo {\em et~al.}(2015)Barredo, Labuhn, Ravets, Lahaye, Browaeys,
  and Adams]{Barredo2015}
Barredo, D., Labuhn, H., Ravets, S., Lahaye, T., Browaeys, A., and Adams, C.
  (2015).
\newblock Coherent excitation transfer in a spin chain of three rydberg atoms.
\newblock {\em Phys. Rev. Lett.}, {\bf 114}, 113002.

\bibitem[Bauer {\em et~al.}(2009)Bauer, Lettner, Vo, Rempe, and
  D\"urr]{Bauer2009}
Bauer, D.~M., Lettner, M., Vo, C., Rempe, G., and D\"urr, S. (2009).
\newblock Combination of a magnetic feshbach resonance and an optical
  bound-to-bound transition.
\newblock {\em Phys. Rev. A\/}, {\bf 79}, 062713.

\bibitem[Baumann {\em et~al.}(2010)Baumann, Guerlin, Brennecke, and
  Esslinger]{Baumann2010}
Baumann, K., Guerlin, C., Brennecke, F., and Esslinger, T. (2010).
\newblock Dicke quantum phase transition with a superfluid gas in an optical
  cavity.
\newblock {\em Nature\/}, {\bf 464}, 1301.

\bibitem[Becker {\em et~al.}(2008)Becker, Stellmer, Soltan-Panahi, D\"orscher,
  Baumert, Richter, Kronj\"ager, Bongs, and Sengstock]{Becker2008}
Becker, C., Stellmer, S., Soltan-Panahi, P., D\"orscher, S., Baumert, M.,
  Richter, E., Kronj\"ager, J., Bongs, K., and Sengstock, K. (2008).
\newblock Oscillations and interactions of dark and dark–bright solitons in
  bose–einstein condensates.
\newblock {\em Nat. Phys.}, {\bf 4}, 496.

\bibitem[Belmechri {\em et~al.}(2013)Belmechri, F\"orster, Alt, Widera,
  Meschede, and Alberti]{Belmechri2013}
Belmechri, N., F\"orster, L., Alt, W., Widera, A., Meschede, D., and Alberti,
  A. (2013).
\newblock Microwave control of atomic motional states in a spin-dependent
  optical lattice.
\newblock {\em New J. Phys.}, {\bf 46}, 104006.

\bibitem[Bergquist {\em et~al.}(1986)Bergquist, Hulet, Itano, and
  Wineland]{Bergquist1986}
Bergquist, J.~C., Hulet, R.~G., Itano, W.~M., and Wineland, D.~J. (1986).
\newblock Observation of quantum jumps in a single atom.
\newblock {\em Phys. Rev. Lett.}, {\bf 57}, 1699.

\bibitem[Birkl {\em et~al.}(1992)Birkl, Kassner, and Walther]{Birkl1992}
Birkl, G., Kassner, S., and Walther, H. (1992).
\newblock Multiple-shell structures of laser-cooled $^{24}$m$^+$ ions in a
  quadrupole storage ring.
\newblock {\em Nature\/}, {\bf 357}, 310.

\bibitem[Blatt and Roos(2012)Blatt and Roos]{Roos2012}
Blatt, R. and Roos, C.~F. (2012).
\newblock Quantum simulation with trapped ions.
\newblock {\em Nature Physics\/}, {\bf 8}, 277.

\bibitem[Blatt {\em et~al.}(2011)Blatt, Nicholson, Bloom, Williams, Thomsen,
  Julienne, and Ye]{Blatt2011}
Blatt, S., Nicholson, T.~L., Bloom, B.~J., Williams, J.~R., Thomsen, J.~W.,
  Julienne, P.~S., and Ye, J. (2011).
\newblock Measurement of optical feshbach resonances in an ideal gas.
\newblock {\em Phys. Rev. Lett.}, {\bf 107}, 073202.

\bibitem[Blaum {\em et~al.}(2010)Blaum, Novikov, and Werth]{Blaum2010}
Blaum, K., Novikov, Y.~N., and Werth, G. (2010).
\newblock Penning traps as a versatile tool for precise experiments in
  fundamental physics.
\newblock {\em Contemp. Phys.}, {\bf 51}, 149.

\bibitem[Bloch(2005)Bloch]{Bloch2005}
Bloch, I. (2005).
\newblock Ultracold quantum gases in optical lattices.
\newblock {\em Nat. Phys.}, {\bf 1}, 23.

\bibitem[Bloch {\em et~al.}(2008)Bloch, Dalibard, and Zwerger]{Bloch_2008}
Bloch, I., Dalibard, J., and Zwerger, W. (2008).
\newblock Many-body physics with ultracold gases.
\newblock {\em Rev. Mod. Phys.}, {\bf 80}, 885.

\bibitem[Bloch {\em et~al.}(2012)Bloch, Dalibard, and Nascimb\`ene]{Bloch2012}
Bloch, I., Dalibard, J., and Nascimb\`ene, S. (2012).
\newblock Quantum simulations with ultracold quantum gases.
\newblock {\em Nat. Phys.}, {\bf 8}, 267.

\bibitem[Bohn and Julienne(1997)Bohn and Julienne]{Bohn1997}
Bohn, J.~L. and Julienne, P.~S. (1997).
\newblock Prospects for influencing scattering lengths with far-off-resonant
  light.
\newblock {\em Phys. Rev. A\/}, {\bf 56}, 1486--1491.

\bibitem[Boozer {\em et~al.}(2006)Boozer, Boca, Miller, Northup, and
  Kimble]{Boozer2006}
Boozer, A.~D., Boca, A., Miller, R., Northup, T.~E., and Kimble, H.~J. (2006).
\newblock Cooling to the ground state of axial motion for one atom strongly
  coupled to an optical cavity.
\newblock {\em Phys. Rev. Lett.}, {\bf 97}, 083602.

\bibitem[Bowman {\em et~al.}(2015)Bowman, Ireland, Bruce, and
  Cassettari]{Bowman2015}
Bowman, D., Ireland, P., Bruce, G.~D., and Cassettari, D. (2015).
\newblock Multi-wavelength holography with a single spatial light modulator for
  ultracold atom experiments.
\newblock {\em Opt. Exp.}, {\bf 23}, 8365.

\bibitem[Bradley {\em et~al.}(1995)Bradley, Sackett, Tollett, and
  Hulet]{Bradley_1995}
Bradley, C.~C., Sackett, C.~A., Tollett, J.~J., and Hulet, R.~G. (1995).
\newblock Evidence of {B}ose-{E}instein condensation in an atomic gas with
  attractive interactions.
\newblock {\em Phys. Rev. Lett.}, {\bf 75}, 1687.

\bibitem[Brandt {\em et~al.}(2011)Brandt, Muldoon, Thiele, Dong, Brainis, and
  Kuhn]{Brandt2010}
Brandt, L., Muldoon, C., Thiele, T., Dong, J., Brainis, E., and Kuhn, A.
  (2011).
\newblock Spatial light modulators for the manipulation of individual atoms.
\newblock {\em Appl. Phys. B\/}, {\bf 102}, 443.

\bibitem[Brantut {\em et~al.}(2012)Brantut, Meineke, Stadler, Krinner, and
  Esslinger]{Brantut_2012}
Brantut, J.-P., Meineke, J., Stadler, D., Krinner, S., and Esslinger, T.
  (2012).
\newblock Conduction of ultracold fermions through a mesoscopic channel.
\newblock {\em Science\/}, {\bf 337}, 1069.

\bibitem[Brazhnyi {\em et~al.}(2009)Brazhnyi, Konotop, P\'erez-Garc\'ia, and
  Ott]{Brazhnyi_2009}
Brazhnyi, V.~A., Konotop, V.~V., P\'erez-Garc\'ia, V.~M., and Ott, H. (2009).
\newblock Dissipation-induced coherent structures in {B}ose-{E}instein
  condensates.
\newblock {\em Phys. Rev. Lett.}, {\bf 102}, 144101.

\bibitem[Brennecke {\em et~al.}(2013)Brennecke, Mottl, Baumann, Landig, Donner,
  and Esslinger]{Brennecke2013}
Brennecke, F., Mottl, R., Baumann, K., Landig, R., Donner, T., and Esslinger,
  T. (2013).
\newblock Real-time observation of fluctuations at the driven-dissipative dicke
  phase transition.
\newblock {\em PNAS\/}, {\bf 110}, 11763.

\bibitem[Breuer and Petruccione(2002)Breuer and Petruccione]{Breuer2002}
Breuer, H.-P. and Petruccione, F. (2002).
\newblock {\em The Theory of Open Quantum Systems\/}.
\newblock Oxford University Press, Oxford, GB.

\bibitem[Britton {\em et~al.}(2012)Britton, Sawyer, Keith, JosephWang,
  Freericks, Uys, Biercuk, and Bollinger]{Britton2012}
Britton, J.~W., Sawyer, B.~C., Keith, A.~C., JosephWang, C.-C., Freericks,
  J.~K., Uys, H., Biercuk, M.~J., and Bollinger, J.~J. (2012).
\newblock Engineered two-dimensional ising interactions in a trapped-ion
  quantum simulator with hundreds of spins.
\newblock {\em Nature\/}, {\bf 484}, 489.

\bibitem[Brown and Gabrielse(1986)Brown and Gabrielse]{Brown1986}
Brown, L.~S. and Gabrielse, G. (1986).
\newblock Geonium theory: Physics of a single electron or ion in a penning
  trap.
\newblock {\em Rev. Mod. Phys.}, {\bf 58}, 233--311.

\bibitem[Carmichael(1993)Carmichael]{Carmichael1993}
Carmichael, H.~J. (1993).
\newblock Quantum trajectory theory for cascaded open systems.
\newblock {\em Phys. Rev. Lett.}, {\bf 70}, 2273.

\bibitem[Cheneau {\em et~al.}(2012)Cheneau, Barmettler, Poletti, Endres,
  Schau\ss{}, Fukuhara, Gro\ss, Bloch, Kollath, and Kuhr]{Cheneau2012}
Cheneau, M., Barmettler, P., Poletti, D., Endres, M., Schau\ss{}, P., Fukuhara,
  T., Gro\ss, C., Bloch, I., Kollath, C., and Kuhr, S. (2012).
\newblock Light-cone-like spreading of correlations in a quantum many-body
  system.
\newblock {\em Nature\/}, {\bf 481}, 484.

\bibitem[Cheuk {\em et~al.}(2015)Cheuk, Nichols, Okan, Gersdorf, Ramasesh,
  Bakr, Lompe, and Zwierlein]{Cheuk2015}
Cheuk, L.~W., Nichols, M.~A., Okan, M., Gersdorf, T., Ramasesh, V.~V., Bakr,
  W.~S., Lompe, T., and Zwierlein, M.~W. (2015).
\newblock Quantum-gas microscope for fermionic atoms.
\newblock {\em Phys. Rev. Lett.}, {\bf 114}, 193001.

\bibitem[Chin {\em et~al.}(2010)Chin, Grimm, Julienne, and Tiesinga]{Chin2010}
Chin, C., Grimm, R., Julienne, P., and Tiesinga, E. (2010).
\newblock Feshbach resonances in ultracold gases.
\newblock {\em Rev. Mod. Phys.}, {\bf 82}, 1225--1286.

\bibitem[Cirac {\em et~al.}(1996)Cirac, Lewenstein, and Zoller]{Cirac1996}
Cirac, J.~I., Lewenstein, M., and Zoller, P. (1996).
\newblock Collective laser cooling of trapped atoms.
\newblock {\em Europhys. Lett.}, {\bf 35}, 9.

\bibitem[Clark {\em et~al.}(2015)Clark, Ha, Xu, and Chin]{Clark2015}
Clark, L.~W., Ha, L.-C., Xu, C.-Y., and Chin, C. (2015).
\newblock Quantum dynamics with spatiotemporal control of interactions in a
  stable bose-einstein condensate.
\newblock {\em arXiv:1506.01766\/}.

\bibitem[Cohen-Tannoudji and Gu\'ery-Odelin(2011)Cohen-Tannoudji and
  Gu\'ery-Odelin]{Cohen2011}
Cohen-Tannoudji, C. and Gu\'ery-Odelin, D. (2011).
\newblock {\em Advances in atomic physics: an overview\/}.
\newblock World Scientific, London, Great Britain.

\bibitem[Cohen-Tannoudji {\em et~al.}(1998)Cohen-Tannoudji, Dupont-Roc, and
  Grynberg]{Cohen1998}
Cohen-Tannoudji, C., Dupont-Roc, J., and Grynberg, G. (1998).
\newblock {\em Atom-Photon interactions\/}.
\newblock WILEY-VCH.

\bibitem[Courtade {\em et~al.}(2004)Courtade, Anderlini, Ciampini, M\"uller,
  Morsch, Arimondo, M.Aymar, and Robinson]{Courtade2004}
Courtade, E., Anderlini, M., Ciampini, D., M\"uller, J.~H., Morsch, O.,
  Arimondo, E., M.Aymar, and Robinson, E.~J. (2004).
\newblock Two-photon ionization of cold rubidium atoms with a near resonant
  intermediate state.
\newblock {\em J. Phys. B\/}, {\bf 37}, 967.

\bibitem[Dagotto(2005)Dagotto]{Dagotto2005}
Dagotto, E. (2005).
\newblock Complexity in strongly correlated electronic systems.
\newblock {\em Science\/}, {\bf 309}, 257.

\bibitem[Dalibard {\em et~al.}(1992)Dalibard, Castin, and
  M\o{}lmer]{Dalibard1992}
Dalibard, J., Castin, Y., and M\o{}lmer, K. (1992).
\newblock Wave-function approach to dissipative processes in quantum optics.
\newblock {\em Phys. Rev. Lett.}, {\bf 68}, 580.

\bibitem[Davis {\em et~al.}(1995)Davis, Mewes, Andrews, van Druten, Durfee,
  Kurn, and Ketterle]{Davis_1995}
Davis, K.~B., Mewes, M.~O., Andrews, M.~R., van Druten, N.~J., Durfee, D.~S.,
  Kurn, D.~M., and Ketterle, W. (1995).
\newblock {B}ose-{E}instein condensation in a gas of sodium atoms.
\newblock {\em Phys. Rev. Lett.}, {\bf 75}, 3969.

\bibitem[Dicke(1954)Dicke]{Dicke1954}
Dicke, R.~H. (1954).
\newblock Coherence in spontaneous radiation processes.
\newblock {\em Phys. Rev.}, {\bf 93}, 99.

\bibitem[Diedrich {\em et~al.}(1989)Diedrich, Bergquist, Itano, and
  Wineland]{Diedrich1989}
Diedrich, F., Bergquist, J.~C., Itano, W.~M., and Wineland, D.~J. (1989).
\newblock Laser cooling to the zero-point energy of motion.
\newblock {\em Phys. Rev. Lett.}, {\bf 62}, 403.

\bibitem[Diehl {\em et~al.}(2008)Diehl, Micheli, Kantian, Kraus, B\"uchlerm,
  and Zoller]{Diehl_2008}
Diehl, S., Micheli, A., Kantian, A., Kraus, B., B\"uchlerm, H.~P., and Zoller,
  P. (2008).
\newblock Quantum states and phases in driven open quantum systems with cold
  atoms.
\newblock {\em Nat. Phys\/}, {\bf 4}, 878.

\bibitem[Diehl {\em et~al.}(2011)Diehl, Rico, Baranov, and Zoller]{Diehl_2011}
Diehl, S., Rico, E., Baranov, M.~A., and Zoller, P. (2011).
\newblock Topology by dissipation in atomic quantum wires.
\newblock {\em Nat. Phys.}, {\bf 7}, 971.

\bibitem[Dodhy {\em et~al.}(1987)Dodhy, Stockdale, Compton, Tang, Lambropoulos,
  and Lyras]{Dodhy1987}
Dodhy, A., Stockdale, J. A.~D., Compton, R.~N., Tang, X., Lambropoulos, P., and
  Lyras, A. (1987).
\newblock Two-photon resonant three-photon ionization of the nd $^{2}d$ states
  of cesium, rubidium, and sodium: Photoelectron angular distributions.
\newblock {\em Phys. Rev. A\/}, {\bf 35}, 2878--2891.

\bibitem[Eckel {\em et~al.}(2014)Eckel, Lee, Jendrzejewski, Murray, Clark,
  Lobb, Phillips, Edwards, and Campbell]{Eckel2014}
Eckel, S., Lee, J.~G., Jendrzejewski, F., Murray, N., Clark, C.~W., Lobb,
  C.~J., Phillips, W.~D., Edwards, M., and Campbell, G.~K. (2014).
\newblock Hysteresis in a quantized superfluid `atomtronic' circuit.
\newblock {\em Nature\/}, {\bf 506}, 200--203.

\bibitem[Edge {\em et~al.}(2015)Edge, Anderson, Jervis, McKay, Day, Trotzky,
  and Thywissen]{Edge2015}
Edge, G. J.~A., Anderson, R., Jervis, D., McKay, D.~C., Day, R., Trotzky, S.,
  and Thywissen, J.~H. (2015).
\newblock Imaging and addressing of individual fermionic atoms in an optical
  lattice.
\newblock {\em arxiv:1510.04744\/}.

\bibitem[Eisert {\em et~al.}(2014)Eisert, Friesdorf, and Gogolin]{Eisert_2014}
Eisert, J., Friesdorf, M., and Gogolin, C. (2014).
\newblock Quantum many-body systems out of equilibrium.
\newblock {\em arXiv:1408.5148\/}.

\bibitem[Endres {\em et~al.}(2011)Endres, Cheneau, Fukuhara, Weitenberg,
  Schau\ss{}, Gro\ss, Mazza, Banuls, Pollet, Bloch, and Kuhr]{Endres2011}
Endres, M., Cheneau, M., Fukuhara, T., Weitenberg, C., Schau\ss{}, P., Gro\ss,
  C., Mazza, L., Banuls, M.~C., Pollet, L., Bloch, I., and Kuhr, S. (2011).
\newblock Observation of correlated particle-hole pairs and string order in
  low-dimensional mott insulators.
\newblock {\em Science\/}, {\bf 334}, 200.

\bibitem[Endres {\em et~al.}(2012)Endres, Fukuhara, Pekker, Cheneau,
  Schau\ss{}, Gro\ss, Demler, Kuhr, and Bloch]{Endres2012}
Endres, M., Fukuhara, T., Pekker, D., Cheneau, M., Schau\ss{}, P., Gro\ss, C.,
  Demler, E., Kuhr, S., and Bloch, I. (2012).
\newblock The ‘higgs’ amplitude mode at the two-dimensional superfluid/mott
  insulator transition.
\newblock {\em Nature\/}, {\bf 487}, 454.

\bibitem[Enomoto {\em et~al.}(2008)Enomoto, Kasa, Kitagawa, and
  Takahashi]{Enomoto2008}
Enomoto, K., Kasa, K., Kitagawa, M., and Takahashi, Y. (2008).
\newblock Optical feshbach resonance using the intercombination transition.
\newblock {\em Phys. Rev. Lett.}, {\bf 101}, 203201.

\bibitem[Eschner {\em et~al.}(2003)Eschner, Morigi, Schmidt-Kaler, and
  Blatt]{Eschner2003}
Eschner, J., Morigi, G., Schmidt-Kaler, F., and Blatt, R. (2003).
\newblock Laser cooling of trapped ions.
\newblock {\em J. Opt. Soc. Am. B\/}, {\bf 20}, 1003.

\bibitem[Fedichev {\em et~al.}(1996)Fedichev, Kagan, Shlyapnikov, and
  Walraven]{Fedichev1996}
Fedichev, P.~O., Kagan, Y., Shlyapnikov, G.~V., and Walraven, J. T.~M. (1996).
\newblock Influence of nearly resonant light on the scattering length in
  low-temperature atomic gases.
\newblock {\em Phys. Rev. Lett.}, {\bf 77}, 2913--2916.

\bibitem[Fischer {\em et~al.}(2001)Fischer, Guti\'errez-Medina, and
  Raizen]{Fischer2001}
Fischer, M.~C., Guti\'errez-Medina, B., and Raizen, M.~G. (2001).
\newblock Observation of the quantum zeno and anti-zeno effects in an unstable
  system.
\newblock {\em Phys. Rev. Lett.}, {\bf 87}, 040402.

\bibitem[F\"olling {\em et~al.}(2005)F\"olling, Gerbier, Widera, Mandel,
  Gericke, and Bloch]{Foelling_2005}
F\"olling, S., Gerbier, F., Widera, A., Mandel, O., Gericke, T., and Bloch, I.
  (2005).
\newblock Spatial quantum noise interferometry in expanding ultracold atom
  clouds.
\newblock {\em Nature\/}, {\bf 434}, 481.

\bibitem[Fort\'{a}gh and Zimmermann(2007)Fort\'{a}gh and
  Zimmermann]{Fortagh2007}
Fort\'{a}gh, J. and Zimmermann, C. (2007).
\newblock Magnetic microtraps for ultracold atoms.
\newblock {\em Rev. Mod. Phys.}, {\bf 79}, 235.

\bibitem[Frese {\em et~al.}(2000)Frese, Ueberholz, Kuhr, Alt, Schrader, Gomer,
  and Meschede]{Frese2000}
Frese, D., Ueberholz, B., Kuhr, S., Alt, W., Schrader, D., Gomer, V., and
  Meschede, D. (2000).
\newblock Single atoms in an optical dipole trap: Towards a deterministic
  source of cold atoms.
\newblock {\em Phys. Rev. Lett.}, {\bf 85}, 3777--3780.

\bibitem[Fu {\em et~al.}(2013)Fu, Wang, Huang, Meng, Hu, and Zhang]{Fu2013}
Fu, Z., Wang, P., Huang, L., Meng, Z., Hu, H., and Zhang, J. (2013).
\newblock Optical control of a magnetic feshbach resonance in an ultracold
  fermi gas.
\newblock {\em Phys. Rev. A\/}, {\bf 88}, 041601.

\bibitem[Fuhrmanek {\em et~al.}(2010)Fuhrmanek, Sortais, Grangier, and
  Browaeys]{Fuhrmanek2010}
Fuhrmanek, A., Sortais, Y., Grangier, P., and Browaeys, A. (2010).
\newblock Measurement of the atom number distribution in an optical tweezer
  using single photon counting.
\newblock {\em Phys. Rev. A\/}, {\bf 82}, 023623.

\bibitem[Fukuhara {\em et~al.}(2013a)Fukuhara, Schau\ss{}, Endres, Hild,
  Cheneau, Bloch, and Gro\ss]{Fukuhara2013a}
Fukuhara, T., Schau\ss{}, P., Endres, M., Hild, S., Cheneau, M., Bloch, I., and
  Gro\ss, C. (2013a).
\newblock Microscopic observation of magnon bound states and their dynamics.
\newblock {\em Nature\/}, {\bf 502}, 76.

\bibitem[Fukuhara {\em et~al.}(2013b)Fukuhara, Kantian, Endres, Cheneau,
  Schau\ss{}, Hild, Bellem, Schollw\"ock, Giamarchi, Gro\ss, Bloch, and
  Kuhr]{Fukuhara2013}
Fukuhara, T., Kantian, A., Endres, M., Cheneau, M., Schau\ss{}, P., Hild, S.,
  Bellem, D., Schollw\"ock, U., Giamarchi, T., Gro\ss, C., Bloch, I., and Kuhr,
  S. (2013b).
\newblock Quantum dynamics of a mobile spin impurity.
\newblock {\em Nature Phys.}, {\bf 9}, 235.

\bibitem[Fukuhara {\em et~al.}(2015)Fukuhara, Hild, Zeiher, Schau\ss, Bloch,
  Endres, and Gro\ss]{Fukuhara2015}
Fukuhara, T., Hild, S., Zeiher, J., Schau\ss, P., Bloch, I., Endres, M., and
  Gro\ss, C. (2015).
\newblock Spatially resolved detection of a spin-entanglement wave in a
  bose-hubbard chain.
\newblock {\em arxiv:1504.02582\/}.

\bibitem[Gabbanini {\em et~al.}(1997)Gabbanini, Gozzini, and
  Lucchesini]{Gabbanini1997}
Gabbanini, C., Gozzini, S., and Lucchesini, A. (1997).
\newblock Photoionization cross section measurement in a rb vapor cell trap.
\newblock {\em Opt. Commun.}, {\bf 141}, 25.

\bibitem[Gallagher(1988)Gallagher]{Gallagher1988}
Gallagher, T.~F. (1988).
\newblock Rydberg atoms.
\newblock {\em Rep. Prog. Phys.}, {\bf 51}, 143.

\bibitem[Gemelke {\em et~al.}(2009)Gemelke, Zhang, Hung, and Chin]{Gemelke2009}
Gemelke, N., Zhang, X., Hung, C.-L., and Chin, C. (2009).
\newblock In situ observation of incompressible mott-insulating domains in
  ultracold atomic gases.
\newblock {\em Nature\/}, {\bf 460}, 995.

\bibitem[Gericke {\em et~al.}(2006)Gericke, Utfeld, Hommerstad, and
  Ott]{Gericke_2006}
Gericke, T., Utfeld, C., Hommerstad, N., and Ott, H. (2006).
\newblock A scanning electron microscope for ultracold atoms.
\newblock {\em Laser Phys. Lett.}, {\bf 3}(8), 415.

\bibitem[Gericke {\em et~al.}(2007)Gericke, W\"urtz, Reitz, Utfeld, and
  Ott]{Gericke_2007}
Gericke, T., W\"urtz, P., Reitz, D., Utfeld, C., and Ott, H. (2007).
\newblock All-optical formation of a {B}ose-{E}instein condensate for
  applications in scanning electron microscopy.
\newblock {\em Appl. Phys. B\/}, {\bf 89}, 447.

\bibitem[Gericke {\em et~al.}(2008)Gericke, W\"urtz, Reitz, Langen, and
  Ott]{Gericke_2008}
Gericke, T., W\"urtz, P., Reitz, D., Langen, T., and Ott, H. (2008).
\newblock High-resolution scanning electron microscopy of an ultracold quantum
  gas.
\newblock {\em Nat. Phys.}, {\bf 4}, 949.

\bibitem[Gleyzes {\em et~al.}(2007)Gleyzes, Kuhr, Guerlin, Bernu,
  Del\'{e}glise, Busk~Ho, Brune, Raimond, and Haroche]{Gleyzes2007}
Gleyzes, S., Kuhr, S., Guerlin, C., Bernu, J., Del\'{e}glise, S., Busk~Ho, U.,
  Brune, M., Raimond, J.~M., and Haroche (2007).
\newblock Quantum jumps of light recording the birth and death of a photon in a
  cavity.
\newblock {\em Nature\/}, {\bf 446}, 297.

\bibitem[Gomer and Meschede(2001)Gomer and Meschede]{Gomer2001}
Gomer, V. and Meschede, D. (2001).
\newblock A single trapped atom: Light-matter interaction at the microscopic
  level.
\newblock {\em Ann. Phys.}, {\bf 10}, 9.

\bibitem[Greiner {\em et~al.}(2002)Greiner, Mandel, Esslinger, Hänsch, and
  Bloch1]{Greiner2002}
Greiner, M., Mandel, O., Esslinger, T., Hänsch, T.~W., and Bloch1, I. (2002).
\newblock Quantum phase transition from a superfluid to a mott insulator in a
  gas of ultracold atoms.
\newblock {\em Nature\/}, {\bf 415}, 39.

\bibitem[Grimm {\em et~al.}(2000)Grimm, Weidem\"uller, and
  Ovchinnikov]{Grimm2000}
Grimm, R., Weidem\"uller, M., and Ovchinnikov, Y.~B. (2000).
\newblock Optical dipole traps for neutral atoms.
\newblock {\em Adv. At. Mol. Opt. Phys.}, {\bf 42}, 95.

\bibitem[Gr\"uner {\em et~al.}(2009)Gr\"uner, Jag, Stibor, Visanescu,
  H\"affner, Kern, G\"unther, and Fort\'{a}gh]{Grüner2009}
Gr\"uner, B., Jag, M., Stibor, A., Visanescu, G., H\"affner, M., Kern, D.,
  G\"unther, A., and Fort\'{a}gh, J. (2009).
\newblock Integrated atom detector based on field ionization near carbon
  nanotubes.
\newblock {\em Phys. Rev. A\/}, {\bf 80}, 063422.

\bibitem[Gr\"unzweig {\em et~al.}(2010)Gr\"unzweig, Hilliard, McGovern, and
  Andersen]{Grünzweig2010}
Gr\"unzweig, T., Hilliard, A., McGovern, M., and Andersen, M.~F. (2010).
\newblock Near-deterministic preparation of a single atom in an optical
  microtrap.
\newblock {\em Nature Phys.}, {\bf 6}, 951.

\bibitem[Grynberg {\em et~al.}(2010)Grynberg, Aspect, and Fabre]{Aspect2010}
Grynberg, G., Aspect, A., and Fabre, C. (2010).
\newblock {\em Introduction to Quantum Optics\/}.
\newblock Cambridge University Press.

\bibitem[Guarrera and Ott(2011)Guarrera and Ott]{Guarrera_2011}
Guarrera, V. and Ott, H. (2011).
\newblock Absolute electron-impact ionization cross section measurements using
  a magneto-optical trap.
\newblock {\em Adv. Imag. Electron Phys.}, {\bf 169}, 75.

\bibitem[Guarrera {\em et~al.}(2011)Guarrera, W\"urtz, Ewerbeck, Vogler,
  Barontini, and Ott]{Guarrera_correlations_2011}
Guarrera, V., W\"urtz, P., Ewerbeck, A., Vogler, A., Barontini, G., and Ott, H.
  (2011).
\newblock Observation of local temporal correlations in trapped quantum gases.
\newblock {\em Phys. Rev. Lett.}, {\bf 107}, 160403.

\bibitem[Guarrera {\em et~al.}(2012)Guarrera, Muth, Labouvie, Vogler,
  Barontini, Fleischhauer, and Ott]{Guarrera_2012}
Guarrera, V., Muth, D., Labouvie, R., Vogler, A., Barontini, G., Fleischhauer,
  M., and Ott, H. (2012).
\newblock Spatiotemporal fermionization of strongly interacting one-dimensional
  bosons.
\newblock {\em Phys. Rev. A\/}, {\bf 86}, 021601.

\bibitem[Haas {\em et~al.}(2014)Haas, Volz, Gehr, Reichel, and
  Est\`{e}ve]{Haas_2014}
Haas, F., Volz, J., Gehr, R., Reichel, J., and Est\`{e}ve, J. (2014).
\newblock Entangled states of more than 40 atoms in an optical fiber cavity.
\newblock {\em Science\/}, {\bf 344}, 180.

\bibitem[Haller {\em et~al.}(2015)Haller, Hudson, Kelly, Cotta, Peaudecerf,
  Bruce, and Kuhr]{Haller2015}
Haller, E., Hudson, J., Kelly, A., Cotta, D.~A., Peaudecerf, B., Bruce, G.~D.,
  and Kuhr, S. (2015).
\newblock Single-atom imaging of fermions in a quantum-gas microscope.
\newblock {\em doi:10.1038/nphys3403\/}.

\bibitem[Haubrich and und D.~Meschede(1993)Haubrich and und
  D.~Meschede]{Haubrich1993}
Haubrich, D. and und D.~Meschede, A.~H. (1993).
\newblock A simple model for optical capture of atoms in strong magnetic
  quadrupole fields.
\newblock {\em Opt. Comm.}, {\bf 102}, 225.

\bibitem[Haubrich {\em et~al.}(1996)Haubrich, Schadwinkel, Strauch, Ueberholz,
  Wynands, and Meschede]{Haubrich1996}
Haubrich, D., Schadwinkel, H., Strauch, F., Ueberholz, B., Wynands, R., and
  Meschede, D. (1996).
\newblock Observation of individual neutral atoms in magnetic and
  magneto-optical traps.
\newblock {\em Europhys. Lett.}, {\bf 34}, 663.

\bibitem[Hild {\em et~al.}(2014)Hild, Fukuhara, Schau\ss{}, Zeiher, Knap,
  Demler, Bloch, and Gro\ss]{Hild2014}
Hild, S., Fukuhara, T., Schau\ss{}, P., Zeiher, J., Knap, M., Demler, E.,
  Bloch, I., and Gro\ss, C. (2014).
\newblock Far-from-equilibrium spin transport in heisenberg quantum magnets.
\newblock {\em Phys. Rev. Lett.}, {\bf 113}, 147205.

\bibitem[Hodgman {\em et~al.}(2011)Hodgman, Dall, Manning, Baldwin, and
  Truscott]{Hodgman2011}
Hodgman, S.~S., Dall, R.~G., Manning, A.~G., Baldwin, K. G.~H., and Truscott,
  A.~G. (2011).
\newblock Direct measurement of long-range third-order coherence in
  bose-einstein condensates.
\newblock {\em Science\/}, {\bf 331}, 1046.

\bibitem[Hood {\em et~al.}(2000)Hood, Lynn, Doherty, Parkins, and
  Kimble]{Hood2000}
Hood, C.~J., Lynn, T.~W., Doherty, A.~C., Parkins, A.~S., and Kimble, H.~J.
  (2000).
\newblock The atom-cavity microscope: Single atoms bound in orbit by single
  photons.
\newblock {\em Science\/}, {\bf 287}, 1457.

\bibitem[Hu and Kimble(1994)Hu and Kimble]{Hu1994}
Hu, Z. and Kimble, H.~J. (1994).
\newblock Observation of a single atom in a magneto-optical trap.
\newblock {\em Opt. Lett.}, {\bf 19}, 1888.

\bibitem[Hume {\em et~al.}(2013)Hume, Stroescu, Joos, Muessel, Strobel, and
  Oberthaler]{Hume2013}
Hume, D.~B., Stroescu, I., Joos, M., Muessel, W., Strobel, H., and Oberthaler,
  M.~K. (2013).
\newblock Accurate atom counting in mesoscopic ensembles.
\newblock {\em Phys. Rev. Lett.}, {\bf 111}, 253001.

\bibitem[Hung {\em et~al.}(2011)Hung, Zhang, Gemelke, and Chin]{Hung2011}
Hung, C.-L., Zhang, X., Gemelke, N., and Chin, C. (2011).
\newblock Observation of scale invariance and universality in two-dimensional
  bose gases.
\newblock {\em Nature\/}, {\bf 470}, 236.

\bibitem[Islam {\em et~al.}(2015)Islam, Ma, Preiss, Tai, Lukin, Rispoli, and
  Greiner]{Islam2015}
Islam, R., Ma, R., Preiss, P.~M., Tai, M.~E., Lukin, A., Rispoli, M., and
  Greiner, M. (2015).
\newblock Measuring entanglement entropy through the interference of quantum
  many-body twins.
\newblock {\em arxiv:1509.01160\/}.

\bibitem[Itah {\em et~al.}(2010)Itah, Veksler, Lahav, Blumkin, Moreno, Gordon,
  and Steinhauer]{Itah2010}
Itah, A., Veksler, H., Lahav, O., Blumkin, A., Moreno, C., Gordon, C., and
  Steinhauer, J. (2010).
\newblock Direct observation of a sub-poissonian number distribution of atoms
  in an optical lattice.
\newblock {\em Phys. Rev. Lett.}, {\bf 104}, 113001.

\bibitem[Itano {\em et~al.}(1990)Itano, Heinzen, Bollinger, and
  Wineland]{Itano1990}
Itano, W.~M., Heinzen, D.~J., Bollinger, J.~J., and Wineland, D.~J. (1990).
\newblock Quantum zeno effect.
\newblock {\em Phys. Rev. A\/}, {\bf 41}, 2295.

\bibitem[Jaynes and Cummings(1963)Jaynes and Cummings]{Jaynes1963}
Jaynes, E.~T. and Cummings, F.~W. (1963).
\newblock Comparison of quantum and semiclassical radiation theories with
  application to the beam maser.
\newblock {\em Proc. IEEE\/}, {\bf 51}, 89.

\bibitem[Jeltes {\em et~al.}(2007)Jeltes, McNamara, Hogervorst, Vassen,
  Krachmalnicoff, Schellekens, Perrin, Chang, Boiron, Aspect, and
  Westbrook]{Jeltes2007}
Jeltes, T., McNamara, J.~M., Hogervorst, W., Vassen, W., Krachmalnicoff, V.,
  Schellekens, M., Perrin, A., Chang, H., Boiron, D., Aspect, A., and
  Westbrook, C.~I. (2007).
\newblock Comparison of the hanbury brown–twiss effect for bosons and
  fermions.
\newblock {\em Nature\/}, {\bf 445}, 402.

\bibitem[Jiang {\em et~al.}(2011)Jiang, Williams, Bailey, Davis, Hu, Lu,
  O’Connor, Purtschert, Sturchio, Sun, and Mueller]{Jiang2011}
Jiang, W., Williams, W., Bailey, K., Davis, A., Hu, S.-M., Lu, Z.-T.,
  O’Connor, T., Purtschert, R., Sturchio, N., Sun, Y., and Mueller, P.
  (2011).
\newblock $^{39}$ar detection at the $10^{-16}$ isotopic abundance level with
  atom trap trace analysis.
\newblock {\em Phys. Rev. Lett.}, {\bf 106}, 103001.

\bibitem[Jurcevic {\em et~al.}(2014)Jurcevic, Lanyon, Hauke, Hempel, Zoller,
  Blatt, and Roos]{Jurcevic2014}
Jurcevic, P., Lanyon, B.~P., Hauke, P., Hempel, C., Zoller, P., Blatt, R., and
  Roos, C.~F. (2014).
\newblock Quasiparticle engineering and entanglement propagation in a quantum
  many-body system.
\newblock {\em Nature\/}, {\bf 511}, 202.

\bibitem[Karski {\em et~al.}(2009)Karski, F\"orster, Choi, Alt, Widera, and
  Meschede]{Karski2009}
Karski, M., F\"orster, L., Choi, J., Alt, W., Widera, A., and Meschede, D.
  (2009).
\newblock Nearest-neighbor detection of atoms in a 1d optical lattice by
  fluorescence imaging.
\newblock {\em Phys. Rev. Lett.}, {\bf 102}, 053001.

\bibitem[Ketterle and Zwierlein(2008)Ketterle and Zwierlein]{Ketterle2008}
Ketterle, W. and Zwierlein, M.~W. (2008).
\newblock Making, probing and understanding ultracold {F}ermi gases.
\newblock {\em Proceedings of the International School of Physics ``Enrico
  Fermi"\/}, {\bf (IOS Press, Amsterdam, 2008)}, 95.

\bibitem[Ketterle {\em et~al.}(1999)Ketterle, Durfee, and
  Stampper-Kurn]{Ketterle1999}
Ketterle, W., Durfee, D., and Stampper-Kurn, D. (1999).
\newblock Making, probing and understanding bose-einstein condensates.
\newblock In {\em Bose-Einstein condensation in atomic gases, Proceedings of
  the International School of Physics "Enrico Fermi", Course CXL, edited by M.
  Inguscio, S. Stringari and C.E. Wieman (IOS Press)\/}, page~67, Amsterdam,
  The Netherlands".

\bibitem[Knoernschild {\em et~al.}(2010)Knoernschild, Zhang, Isenhower, Gill,
  Lu, Saffman, and Kim]{Knoernschild2010}
Knoernschild, C., Zhang, L., Isenhower, L., Gill, A.~T., Lu, F.~P., Saffman,
  M., and Kim, J. (2010).
\newblock Independent individual addressing of multiple neutral atom qubits
  with a micromirror-based beam steering system.
\newblock {\em Appl. Phys. Lett.}, {\bf 97}, 134101.

\bibitem[Koizumi and Chihara(2009)Koizumi and Chihara]{Koizumi2008}
Koizumi, T. and Chihara, Y. (2009).
\newblock Absolute detection efficiencies of an ion-counting system with a
  channel-electron multiplier.
\newblock {\em J. Phys.: Conf. Ser.}, {\bf 163}, 012114.

\bibitem[Kraft {\em et~al.}(2007)Kraft, G\"unther, Fort\'agh, and
  Zimmermann]{Kraft2007}
Kraft, S., G\"unther, A., Fort\'agh, J., and Zimmermann, C. (2007).
\newblock Spatially resolved photoionization of ultracold atoms on an atom
  chip.
\newblock {\em Phys. Rev. A\/}, {\bf 75}, 063605.

\bibitem[Kuhr {\em et~al.}(2001)Kuhr, Alt, Schrader, M\"uller, Gomer, and
  Meschede]{Kuhr2001}
Kuhr, S., Alt, W., Schrader, D., M\"uller, M., Gomer, V., and Meschede, D.
  (2001).
\newblock Deterministic delivery of a single atom.
\newblock {\em Science\/}, {\bf 293}, 278.

\bibitem[Labouvie {\em et~al.}(2015a)Labouvie, Heun, Santra, and
  Ott]{Labouvie2015b}
Labouvie, R., Heun, S., Santra, B., and Ott, H. (2015a).
\newblock Bistability in a driven-dissipative superfluid.
\newblock {\em arXiv:1507.05007\/}.

\bibitem[Labouvie {\em et~al.}(2015b)Labouvie, Santra, Heun, Wimberger, and
  Ott]{Labouvie2015}
Labouvie, R., Santra, B., Heun, S., Wimberger, S., and Ott, H. (2015b).
\newblock Negative differential conductivity in an interacting quantum gas.
\newblock {\em Phys. Rev. Lett.}, {\bf 115}, 050601.

\bibitem[Langen {\em et~al.}(2014)Langen, Geiger, and
  Schmiedmayer]{Langen_2014}
Langen, T., Geiger, R., and Schmiedmayer, J. (2014).
\newblock Ultracold atoms out of equilibrium.
\newblock {\em arXiv:1408.6377\/}.

\bibitem[Lester {\em et~al.}(2014)Lester, Kaufman, and Regal]{Lester2014}
Lester, B.~J., Kaufman, A.~M., and Regal, C.~A. (2014).
\newblock Raman cooling imaging: Detecting single atoms near their ground state
  of motion.
\newblock {\em Phys. Rev. A\/}, {\bf 90}, 011804.

\bibitem[Letokhov {\em et~al.}(1995)Letokhov, Olshanii, and R.]{Lethokov1995}
Letokhov, V.~S., Olshanii, M.~A., and R., O. Y.~B. (1995).
\newblock Laser cooling of atoms: a review.
\newblock {\em Quantum Semiclass. Opt.}, {\bf 7}, 5.

\bibitem[Lieb and Robinson(1972)Lieb and Robinson]{Lieb1972}
Lieb, E.~H. and Robinson, D.~W. (1972).
\newblock The finite group velocity of quantum spin systems.
\newblock {\em Commun. Math. Phys.}, {\bf 28}, 251.

\bibitem[Lin {\em et~al.}(2009)Lin, Compton, Jim\`enez-Garc\`ia, Porto, and
  Spielman]{Lin_2009}
Lin, Y.-J., Compton, R.~L., Jim\`enez-Garc\`ia, K., Porto, J.~V., and Spielman,
  I.~B. (2009).
\newblock Synthetic magnetic fields for ultracold neutral atoms.
\newblock {\em Nature\/}, {\bf 462}, 628.

\bibitem[Lin {\em et~al.}(2011)Lin, Compton, Jim\`enez-Garc\`ia, Phillips,
  Porto, and Spielman]{Lin_2011}
Lin, Y.-J., Compton, R.~L., Jim\`enez-Garc\`ia, K., Phillips, W.~D., Porto,
  J.~V., and Spielman, I.~B. (2011).
\newblock A synthetic electric force acting on neutral atoms.
\newblock {\em Nat. Phys.}, {\bf 7}, 531.

\bibitem[Loh {\em et~al.}(1990)Loh, Gubernatis, Scalettar, White, Scalapino,
  and Sugar]{Loh1990}
Loh, E.~Y., Gubernatis, J.~E., Scalettar, R.~T., White, S.~R., Scalapino,
  D.~J., and Sugar, R.~L. (1990).
\newblock Sign problem in the numerical simulation of many-electron systems.
\newblock {\em Phys. Rev. B\/}, {\bf 41}, 9301--9307.

\bibitem[Lopes {\em et~al.}(2015)Lopes, Imanaliev, Aspect, Cheneau, Boiron, and
  Westbrook]{Lopes2015}
Lopes, R., Imanaliev, A., Aspect, A., Cheneau, M., Boiron, D., and Westbrook,
  C.~I. (2015).
\newblock Atomic hong-ou-mandel experiment.
\newblock {\em Nature\/}, {\bf 520}, 66.

\bibitem[Lowell {\em et~al.}(2002)Lowell, Northup, Patterson, Takekoshi, and
  Knize]{Lowell2002}
Lowell, J.~R., Northup, T., Patterson, B.~M., Takekoshi, T., and Knize, R.~J.
  (2002).
\newblock Measurement of the photoionization cross section of the $5s_{1/2}$
  state of rubidium.
\newblock {\em Phys. Rev. A\/}, {\bf 66}, 062704.

\bibitem[M.~Saffman and M\o{}lmer(2010)M.~Saffman and M\o{}lmer]{Saffman2010}
M.~Saffman, T. G.~W. and M\o{}lmer, K. (2010).
\newblock Quantum information with rydberg atoms.
\newblock {\em Rev. Mod. Phys.}, {\bf 82}, 2313.

\bibitem[Ma {\em et~al.}(2011)Ma, Tai, Preiss, Bakr, Simon, and
  Greiner]{Ma2011}
Ma, R., Tai, M.~E., Preiss, P., Bakr, W.~S., Simon, J., and Greiner, M. (2011).
\newblock Photon-assisted tunneling in a biased strongly correlated bose gas.
\newblock {\em Phys. Rev. Lett.}, {\bf 107}, 095301.

\bibitem[Madison {\em et~al.}(2000)Madison, Chevy, Wohlleben, and
  Dalibard]{Madison2000}
Madison, K.~W., Chevy, F., Wohlleben, W., and Dalibard, J. (2000).
\newblock Vortex formation in a stirred bose-einstein condensate.
\newblock {\em Phys. Rev. Lett.}, {\bf 84}, 806--809.

\bibitem[Mandel {\em et~al.}(2003)Mandel, Greiner, Widera, Rom, H\"ansch, and
  Bloch]{Mandel2003}
Mandel, O., Greiner, M., Widera, A., Rom, T., H\"ansch, T.~W., and Bloch, I.
  (2003).
\newblock Controlled collisions for multi-particle entanglement of optically
  trapped atoms.
\newblock {\em Nature\/}, {\bf 425}, 937.

\bibitem[Manthey {\em et~al.}(2014)Manthey, Weber, Niederpr\"um, Langer,
  Guarrera, Barontini, and Ott]{Manthey_2014}
Manthey, T., Weber, T.~M., Niederpr\"um, T., Langer, P., Guarrera, V.,
  Barontini, G., and Ott, H. (2014).
\newblock Scanning electron microscopy of {R}ydberg-excited {B}ose-{E}instein
  condensates.
\newblock {\em New J. Phys.}, {\bf 16}, 083034.

\bibitem[Marcassa and Shaffer(2014)Marcassa and Shaffer]{Marcassa2014}
Marcassa, L.~G. and Shaffer, J.~P. (2014).
\newblock Interactions in ultracold rydberg gases.
\newblock {\em Ad. At. Mol. Opt. Phys.}, {\bf 63}, 47.

\bibitem[McConnell {\em et~al.}(2015)McConnell, Zhang, Hu, Cuk, and
  Vuleti\'{c}]{McConnell2015}
McConnell, R., Zhang, H., Hu, J., Cuk, S., and Vuleti\'{c}, V. (2015).
\newblock Entanglement with negative wigner function of almost 3,000 atoms
  heralded by one photon.
\newblock {\em Nature\/}, {\bf 519}, 439.

\bibitem[McKeever {\em et~al.}(2003)McKeever, Buck, Boozer, Kuzmich, Naegerl,
  Stamper-Kurn, and Kimble]{McKeever2003}
McKeever, J., Buck, J.~R., Boozer, A.~D., Kuzmich, A., Naegerl, H.-C.,
  Stamper-Kurn, D.~M., and Kimble, H.~J. (2003).
\newblock State-insensitive cooling and trapping of single atoms in an optical
  cavity.
\newblock {\em Phys. Rev. Lett.}, {\bf 90}, 133602.

\bibitem[Meschede {\em et~al.}(1985)Meschede, Walther, and
  Müller]{Meschede1985}
Meschede, D., Walther, H., and Müller, G. (1985).
\newblock One-atom maser.
\newblock {\em Phys. Rev. Lett.}, {\bf 54}, 551.

\bibitem[Miranda {\em et~al.}(2015)Miranda, Inoue, Okuyama, Nakamoto, and
  Kozuma]{Miranda2015}
Miranda, M., Inoue, R., Okuyama, Y., Nakamoto, A., and Kozuma, M. (2015).
\newblock Site-resolved imaging of ytterbium atoms in a two-dimensional optical
  lattice.
\newblock {\em Phys. Rev. A\/}, {\bf 91}, 063414.

\bibitem[Miroshnychenko {\em et~al.}(2006)Miroshnychenko, Alt, Dotsenko,
  Förster, Khudaverdyan1, Meschede, Schrader, and
  Rauschenbeutel1]{Miroshnychenko2006}
Miroshnychenko, Y., Alt, W., Dotsenko, I., Förster, L., Khudaverdyan1, M.,
  Meschede, D., Schrader, D., and Rauschenbeutel1, A. (2006).
\newblock Quantum engineering: An atom-sorting machine.
\newblock {\em Nature\/}, {\bf 442}, 151.

\bibitem[Monroe {\em et~al.}(1995)Monroe, Meekhof, King, Itano, and
  Wineland]{Monroe1995}
Monroe, C., Meekhof, D.~M., King, B.~E., Itano, W.~M., and Wineland, D.~J.
  (1995).
\newblock Demonstration of a fundamental quantum logic gate.
\newblock {\em Phys. Rev. Lett.}, {\bf 75}, 4714.

\bibitem[Morigi {\em et~al.}(2000)Morigi, Eschner, and Keitel]{Morigi2000}
Morigi, G., Eschner, J., and Keitel, C.~H. (2000).
\newblock Ground state laser cooling using electromagnetically induced
  transparency.
\newblock {\em Phys. Rev. Lett.}, {\bf 85}, 4458--4461.

\bibitem[Nelson {\em et~al.}(2007)Nelson, Li, and Weiss]{Nelson_2007}
Nelson, K.~D., Li, X., and Weiss, D.~S. (2007).
\newblock Imaging single atoms in a three-dimensional array.
\newblock {\em Nat. Phys.}, {\bf 3}, 556.

\bibitem[Neuhauser {\em et~al.}(1980)Neuhauser, Hohenstatt, Toschek, and
  Dehmelt]{Neuhauser1980}
Neuhauser, W., Hohenstatt, M., Toschek, P.~E., and Dehmelt, H. (1980).
\newblock Localized visible ba$+$ mono-ion oscillator.
\newblock {\em Phys. Rev. A\/}, {\bf 22}, 1137.

\bibitem[Nogrette {\em et~al.}(2014)Nogrette, Labuhn, Ravets, Barredo,
  B\'{e}guin, amd T.~Lahaye, and Browaeys]{Nogrette2014}
Nogrette, F., Labuhn, H., Ravets, S., Barredo, D., B\'{e}guin, L., amd
  T.~Lahaye, A.~V., and Browaeys, A. (2014).
\newblock Single atom trapping in holographic 2d arrays of microtraps with
  arbitrary geometries.
\newblock {\em Phys. Rev. X\/}, {\bf 4}, 021034.

\bibitem[Nogues {\em et~al.}(1999)Nogues, Rauschenbeutel, Osnaghi, Brune,
  Raimond, and Haroche]{Nogues1999}
Nogues, G., Rauschenbeutel, A., Osnaghi, S., Brune, M., Raimond, J.~M., and
  Haroche, S. (1999).
\newblock Seeing a single photon without destroying it.
\newblock {\em Nature\/}, {\bf 400}, 239.

\bibitem[Omran {\em et~al.}(2015)Omran, Boll, Hilker, Kleinlein, Salomon,
  Bloch, and Gross]{Omran2015}
Omran, A., Boll, M., Hilker, T., Kleinlein, K., Salomon, G., Bloch, I., and
  Gross, C. (2015).
\newblock Microscopic observation of pauli blocking in degenerate fermionic
  lattice gases.
\newblock {\em arxiv:1510.04599\/}.

\bibitem[\"Ottl {\em et~al.}(2005)\"Ottl, Ritter, K\"ohl, and
  Esslinger]{Öttl2005}
\"Ottl, A., Ritter, S., K\"ohl, M., and Esslinger, T. (2005).
\newblock Correlations and counting statistics of an atom laser.
\newblock {\em Phys. Rev. Lett.}, {\bf 95}, 090404.

\bibitem[Parsons {\em et~al.}(2015)Parsons, Huber, Mazurenko, Chiu, Setiawan,
  Wooley-Brown, Blatt, and Greiner]{Parsons2015}
Parsons, M.~F., Huber, F., Mazurenko, A., Chiu, C.~S., Setiawan, W.,
  Wooley-Brown, K., Blatt, S., and Greiner, M. (2015).
\newblock Site-resolved imaging of fermionic $^{6}\mathrm{Li}$ in an optical
  lattice.
\newblock {\em Phys. Rev. Lett.}, {\bf 114}, 213002.

\bibitem[Patil {\em et~al.}(2014a)Patil, Chakram, Aycock, and
  Vengalattore]{Patil2014}
Patil, Y.~S., Chakram, S., Aycock, L.~M., and Vengalattore, M. (2014a).
\newblock Nondestructive imaging of an ultracold lattice gas.
\newblock {\em Phys. Rev. A\/}, {\bf 90}, 033422.

\bibitem[Patil {\em et~al.}(2014b)Patil, Chakram, and Vengalattore]{Patil2015}
Patil, Y.~S., Chakram, S., and Vengalattore, M. (2014b).
\newblock Quantum control by imaging: The zeno effect in an ultracold lattice
  gas.
\newblock {\em arXiv:1411.2678\/}.

\bibitem[Pepino {\em et~al.}(2009)Pepino, Cooper, Anderson, and
  Holland]{Pepino2009}
Pepino, R.~A., Cooper, J., Anderson, D.~Z., and Holland, M.~J. (2009).
\newblock Atomtronic circuits of diodes and transistors.
\newblock {\em Phys. Rev. Lett.}, {\bf 103}, 140405.

\bibitem[Perrin {\em et~al.}(2007)Perrin, Chang, Krachmalnicoff, Schellekens,
  Boiron, Aspect, and Westbrook]{Perrin2007}
Perrin, A., Chang, H., Krachmalnicoff, V., Schellekens, M., Boiron, D., Aspect,
  A., and Westbrook, C.~I. (2007).
\newblock Observation of atom pairs in spontaneous four-wave mixing of two
  colliding bose-einstein condensates.
\newblock {\em Phys. Rev. Lett.}, {\bf 99}, 150405.

\bibitem[Perrin {\em et~al.}(2012)Perrin, B\"ucker, Manz, Betz, Koller,
  Plisson, Schumm, and Schmiedmayer]{Perrin2012}
Perrin, A., B\"ucker, R., Manz, S., Betz, T., Koller, C., Plisson, T., Schumm,
  T., and Schmiedmayer, J. (2012).
\newblock Hanbury brown and twiss correlations across the bose–einstein
  condensation threshold.
\newblock {\em Nat. Phys.}, {\bf 8}, 195.

\bibitem[Phillips(1998)Phillips]{Phillips1998}
Phillips, W.~D. (1998).
\newblock Laser cooling and trapping of neutral atoms.
\newblock {\em Rev. Mod. Phys.}, {\bf 70}, 721.

\bibitem[Polkovnikov {\em et~al.}(2011)Polkovnikov, Sengupta, Silva, and
  Vengalattore]{Polkovnikov_2011}
Polkovnikov, A., Sengupta, K., Silva, A., and Vengalattore, M. (2011).
\newblock Colloquium: Nonequilibrium dynamics of closed interacting quantum
  systems.
\newblock {\em Rev. Mod. Phys.}, {\bf 83}, 863.

\bibitem[Preiss {\em et~al.}(2015a)Preiss, R.Ma, Tai, Lukin, Rispoli, Y.Lahini,
  Islam, and Greiner]{Preiss2015}
Preiss, P., R.Ma, Tai, M., Lukin, A., Rispoli, M., Y.Lahini, Islam, R., and
  Greiner, M. (2015a).
\newblock Strongly correlated quantum walks in optical lattices.
\newblock {\em Science\/}, {\bf 347}, 1229.

\bibitem[Preiss {\em et~al.}(2015b)Preiss, Ma, Tai, Simon, and
  Greiner]{Preiss2015b}
Preiss, P.~M., Ma, R., Tai, M.~E., Simon, J., and Greiner, M. (2015b).
\newblock Quantum gas microscopy with spin, atom-number, and multilayer
  readout.
\newblock {\em Phys. Rev. A\/}, {\bf 91}, 041602.

\bibitem[Puppe {\em et~al.}(2007)Puppe, Schuster, Grothe, Kubanek, Murr,
  Pinkse, and Rempe]{Puppe2007}
Puppe, T., Schuster, I., Grothe, A., Kubanek, A., Murr, K., Pinkse, P. W.~H.,
  and Rempe, G. (2007).
\newblock Trapping and observing single atoms in a blue-detuned intracavity
  dipole trap.
\newblock {\em Phys. Rev. Lett.}, {\bf 99}, 013002.

\bibitem[Raab {\em et~al.}(1987)Raab, Prentiss, Cable, Steven~Chu, and
  Pritchard]{Raab1987}
Raab, E.~L., Prentiss, M., Cable, A., Steven~Chu, S., and Pritchard, d.~E.
  (1987).
\newblock Trapping of neutral sodium atoms with radiation pressure.
\newblock {\em Phys. Rev. Lett.}, {\bf 59}, 2631.

\bibitem[Randeria and Taylor(2014)Randeria and Taylor]{Randeria2014}
Randeria, M. and Taylor, E. (2014).
\newblock Crossover from bardeen-cooper-schrieffer to bose-einstein
  condensation and the unitary fermi gas.
\newblock {\em Annu. Rev. Cond. Mat. Phys.}, {\bf 5}, 209.

\bibitem[Raussendorf and Briegel(2001)Raussendorf and
  Briegel]{Raussendorf_2001}
Raussendorf, R. and Briegel, H.~J. (2001).
\newblock A one-way quantum computer.
\newblock {\em Phys. Rev. Lett.}, {\bf 86}, 5188.

\bibitem[Reiserer {\em et~al.}(2013)Reiserer, N\"olleke, Ritter, and
  Rempe]{Reiserer2013}
Reiserer, A., N\"olleke, C., Ritter, S., and Rempe, G. (2013).
\newblock Ground-state cooling of a single atom at the center of an optical
  cavity.
\newblock {\em Phys. Rev. Lett.}, {\bf 110}, 223003.

\bibitem[Ritsch {\em et~al.}(2013)Ritsch, Domokos, Brennecke, and
  Esslinger]{Ritsch2013}
Ritsch, H., Domokos, P., Brennecke, F., and Esslinger, T. (2013).
\newblock Cold atoms in cavity-generated dynamical optical potentials.
\newblock {\em Rev. Mod. Phys.}, {\bf 85}, 553.

\bibitem[Ritterbusch {\em et~al.}(2014)Ritterbusch, Ebser, Welte, Reichel,
  Kersting, Purtschert, Aeschbach-Hertig, and Oberthaler]{Ritterbusch2014}
Ritterbusch, F., Ebser, S., Welte, J., Reichel, T., Kersting, A., Purtschert,
  R., Aeschbach-Hertig, W., and Oberthaler, M.~K. (2014).
\newblock Groundwater dating with atom trap trace analysis of $^{39}$ar.
\newblock {\em Geophys. Res. Lett.}, {\bf 41}, 6758.

\bibitem[Rom {\em et~al.}(2007)Rom, Best, van Oosten, Schneider, F\"olling,
  Paredes, and Bloch]{Rom2007}
Rom, T., Best, T., van Oosten, D., Schneider, U., F\"olling, S., Paredes, B.,
  and Bloch, I. (2007).
\newblock Free fermion antibunching in a degenerate atomic fermi gas released
  from an optical lattice.
\newblock {\em Nature\/}, {\bf 444}, 733.

\bibitem[Ronzheimer {\em et~al.}(2013)Ronzheimer, Schreiber, Braun, Hodgman,
  Langer, McCulloch, Heidrich-Meisner, Bloch, and Schneider]{Ronzheimer_2013}
Ronzheimer, J.~P., Schreiber, M., Braun, S., Hodgman, S.~S., Langer, S.,
  McCulloch, I.~P., Heidrich-Meisner, F., Bloch, I., and Schneider, U. (2013).
\newblock Expansion dynamics of interacting bosons in homogeneous lattices in
  one and two dimensions.
\newblock {\em Phys. Rev. Lett.}, {\bf 110}, 205301.

\bibitem[Roos {\em et~al.}(2000)Roos, Leibfried, Mundt, Schmidt-Kaler, Eschner,
  and Blatt]{Roos2000}
Roos, C.~F., Leibfried, D., Mundt, A., Schmidt-Kaler, F., Eschner, J., and
  Blatt, R. (2000).
\newblock Experimental demonstration of ground state laser cooling with
  electromagnetically induced transparency.
\newblock {\em Phys. Rev. Lett.}, {\bf 85}, 5547--5550.

\bibitem[Sachdev {\em et~al.}(2002)Sachdev, Sengupta, and Girvin]{Sachdev2002}
Sachdev, S., Sengupta, K., and Girvin, S.~M. (2002).
\newblock Mott insulators in strong electric fields.
\newblock {\em Phys. Rev. B\/}, {\bf 66}, 075128.

\bibitem[Santos {\em et~al.}(2001)Santos, Léonard, Wang, Barrelet, Perales,
  Rasel, Unnikrishnan, Leduc, and Cohen-Tannoudji]{Santos2001}
Santos, F. P.~D., Léonard, J., Wang, J., Barrelet, C.~J., Perales, F., Rasel,
  E., Unnikrishnan, C.~S., Leduc, M., and Cohen-Tannoudji, C. (2001).
\newblock Bose-einstein condensation of metastable helium.
\newblock {\em Phys. Rev. Lett.}, {\bf 86}, 3459.

\bibitem[Santra and Ott(2015)Santra and Ott]{Santra2015}
Santra, B. and Ott, H. (2015).
\newblock Scanning electron microscopy of cold gases.
\newblock {\em J. Phys. B\/}, {\bf 00}, 0000.

\bibitem[Sauter {\em et~al.}(1986)Sauter, Blatt, Neuhauser, and
  Toschek]{Sauter1986}
Sauter, T., Blatt, R., Neuhauser, W., and Toschek, P.~E. (1986).
\newblock 'quantum jumps' observed in the fluorescence of a single ion.
\newblock {\em Opt. Communic.}, {\bf 60}, 287.

\bibitem[Schau\ss{} {\em et~al.}(2015)Schau\ss{}, Zeiher, Fukuhara, Hild,
  Cheneau, Macri, Pohl, Bloch, and Gro\ss]{Schauss2015}
Schau\ss{}, P., Zeiher, J., Fukuhara, T., Hild, S., Cheneau, M., Macri, T.,
  Pohl, T., Bloch, I., and Gro\ss, C. (2015).
\newblock Crystallization in ising quantum magnets.
\newblock {\em Science\/}, {\bf 347}, 1455.

\bibitem[Schellekens {\em et~al.}(2005)Schellekens, Hoppeler, Perrin, Gomes,
  Boiron, Aspect, and Westbrook]{Schellekens2005}
Schellekens, M., Hoppeler, R., Perrin, A., Gomes, J.~V., Boiron, D., Aspect,
  A., and Westbrook, C.~I. (2005).
\newblock Hanbury brown twiss effect for ultracold quantum gases.
\newblock {\em Science\/}, {\bf 310}, 648.

\bibitem[Schlosser {\em et~al.}(2001)Schlosser, Reymond, Protsenko, and
  Grangier]{Schlosser2001}
Schlosser, N., Reymond, G., Protsenko, I., and Grangier, P. (2001).
\newblock Sub-poissonian loading of single atoms in a microscopic dipole trap.
\newblock {\em Nature\/}, {\bf 411}, 1024.

\bibitem[Schmidt-Kaler {\em et~al.}(2003)Schmidt-Kaler, H\"affner, Riebe,
  Gulde, Lancaster, Deuschle, Becher, Roos, Eschner, and
  Blatt]{SchmidtKaler2003}
Schmidt-Kaler, F., H\"affner, H., Riebe, M., Gulde, S., Lancaster, G. P.~T.,
  Deuschle, T., Becher, C., Roos, C., Eschner, J., and Blatt, J. (2003).
\newblock Realization of the cirac–zoller controlled-not quantum gate.
\newblock {\em Nature\/}, {\bf 422}, 408.

\bibitem[Schrader {\em et~al.}(2004)Schrader, Dotsenko, Khudaverdyan,
  Miroshnychenko, Rauschenbeutel, and Meschede]{Schrader2004}
Schrader, D., Dotsenko, I., Khudaverdyan, M., Miroshnychenko, Y.,
  Rauschenbeutel, A., and Meschede, D. (2004).
\newblock Neutral atom quantum register.
\newblock {\em Phys. Rev. Lett.}, {\bf 93}, 150501.

\bibitem[Schr\"odinger(1952)Schr\"odinger]{Schrödinger1952}
Schr\"odinger, E. (1952).
\newblock Are there quantum jumps? part ii.
\newblock {\em Br. J. Phil. Sci.}, {\bf 3}, 233.

\bibitem[Schwarzkopf {\em et~al.}(2011)Schwarzkopf, Sapiro, and
  Raithel]{Schwarzkopf2011}
Schwarzkopf, A., Sapiro, R.~E., and Raithel, G. (2011).
\newblock Imaging spatial correlations of {R}ydberg excitations in cold atom
  clouds.
\newblock {\em Phys. Rev. Lett.}, {\bf 107}, 103001.

\bibitem[Schwarzkopf {\em et~al.}(2013)Schwarzkopf, Anderson, Thaicharoen, and
  Raithel]{Schwarzkopf2013}
Schwarzkopf, A., Anderson, D.~A., Thaicharoen, N., and Raithel, G. (2013).
\newblock Spatial correlations between rydberg atoms in an optical dipole trap.
\newblock {\em Phys. Rev. A\/}, {\bf 88}, 061406(R).

\bibitem[Serwane {\em et~al.}(2011)Serwane, Z\"urn, Lompe, Ottenstein, Wenz,
  and Jochim]{Serwane2011}
Serwane, F., Z\"urn, G., Lompe, T., Ottenstein, T.~B., Wenz, A.~N., and Jochim,
  S. (2011).
\newblock Deterministic preparation of a tunable few-fermion system.
\newblock {\em Science\/}, {\bf 332}, 336.

\bibitem[Sherson {\em et~al.}(2010)Sherson, Weitenberg, Endres, Cheneau, Bloch,
  and Kuhr]{Sherson_2010}
Sherson, J.~F., Weitenberg, C., Endres, M., Cheneau, M., Bloch, I., and Kuhr,
  S. (2010).
\newblock Single-atom-resolved fluorescence imaging of an atomic {M}ott
  insulator.
\newblock {\em Nature\/}, {\bf 467}, 68.

\bibitem[Shotter(2011)Shotter]{Shotter2011}
Shotter, M.~D. (2011).
\newblock Large-photon-number extraction from individual atoms trapped in an
  optical lattice.
\newblock {\em Phys. Rev. A\/}, {\bf 83}, 033420.

\bibitem[Simon {\em et~al.}(2011)Simon, Bakr, Ma, Tai, Preiss, and
  Greiner]{Simon2011}
Simon, J., Bakr, W.~S., Ma, R., Tai, M.~E., Preiss, P.~M., and Greiner, M.
  (2011).
\newblock Quantum simulation of antiferromagnetic spin chains in an optical
  lattice.
\newblock {\em Nature\/}, {\bf 472}, 307.

\bibitem[Sompet {\em et~al.}(2013)Sompet, Carpentier, Fung, McGovern, and
  Andersen]{Sompet2013}
Sompet, P., Carpentier, A.~V., Fung, Y.~H., McGovern, M., and Andersen, M.~F.
  (2013).
\newblock Dynamics of two atoms undergoing light-assisted collisions in an
  optical microtrap.
\newblock {\em Phys. Rev. A\/}, {\bf 88}, 051401.

\bibitem[Spethmann {\em et~al.}(2012)Spethmann, Kindermann, John, Weber,
  Meschede, and Widera]{Spethmann2012}
Spethmann, N., Kindermann, F., John, S., Weber, C., Meschede, D., and Widera,
  A. (2012).
\newblock Dynamics of single neutral impurity atoms immersed in an ultracold
  gas.
\newblock {\em Phys. Rev. Lett.}, {\bf 109}, 235301.

\bibitem[Stamper-Kurn and Ueda(2013)Stamper-Kurn and Ueda]{Stamper-Kurn_2013}
Stamper-Kurn, D.~M. and Ueda, M. (2013).
\newblock Spinor {B}ose gases: Symmetries, magnetism, and quantum dynamics.
\newblock {\em Rev. Mod. Phys.}, {\bf 85}, 1191.

\bibitem[Stenholm(1986)Stenholm]{Stenholm1986}
Stenholm, S. (1986).
\newblock The semiclassical theory of laser cooling.
\newblock {\em Rev. Mod. Phys.}, {\bf 58}, 699.

\bibitem[Stibor {\em et~al.}(2010)Stibor, Bender, K\"uhnhold, Fort\'{a}gh,
  Zimmermann, and G\"unther]{Stibor2010}
Stibor, A., Bender, H., K\"uhnhold, S., Fort\'{a}gh, J., Zimmermann, C., and
  G\"unther, A. (2010).
\newblock Single-atom detection on a chip: from realization to application.
\newblock {\em New J. Phys.}, {\bf 12}, 065034.

\bibitem[Tavis and Cummings(1968)Tavis and Cummings]{Tavis1968}
Tavis, M. and Cummings, F.~W. (1968).
\newblock Exact solution for an $n$-molecule—radiation-field hamiltonian.
\newblock {\em Phys. Rev.}, {\bf 170}, 379.

\bibitem[Tey {\em et~al.}(2008)Tey, Chen, Aljunid, Chng, Huber, Maslennikov,
  and Kurtsiefer]{Tey2008}
Tey, M.~K., Chen, Z., Aljunid, S.~A., Chng, B., Huber, F., Maslennikov, G., and
  Kurtsiefer, C. (2008).
\newblock Strong interaction between light and a single trapped atom without
  the need for a cavity.
\newblock {\em Nat. Phys.}, {\bf 4}, 924.

\bibitem[Theis {\em et~al.}(2004)Theis, Thalhammer, Winkler, Hellwig, Ruff,
  Grimm, and Denschlag]{Theis2004}
Theis, M., Thalhammer, G., Winkler, K., Hellwig, M., Ruff, G., Grimm, R., and
  Denschlag, J.~H. (2004).
\newblock Tuning the scattering length with an optically induced feshbach
  resonance.
\newblock {\em Phys. Rev. Lett.}, {\bf 93}, 123001.

\bibitem[Tychkov {\em et~al.}(2006)Tychkov, Jeltes, McNamara, Tol, Herschbach,
  Hogervorst, and Vassen]{Tychkov2006}
Tychkov, A.~S., Jeltes, T., McNamara, J.~M., Tol, P. J.~J., Herschbach, N.,
  Hogervorst, W., and Vassen, W. (2006).
\newblock Metastable helium bose-einstein condensate with a large number of
  atoms.
\newblock {\em Phys. Rev. A\/}, {\bf 73}, 031603.

\bibitem[Vassen {\em et~al.}(2012)Vassen, Cohen-Tannoudji, Leduc, Boiron,
  Westbrook, Truscott, Baldwin, Birkl, Cancio, and Trippenbach]{Vassen2012}
Vassen, W., Cohen-Tannoudji, C., Leduc, M., Boiron, D., Westbrook, C.~I.,
  Truscott, A., Baldwin, K., Birkl, G., Cancio, P., and Trippenbach, M. (2012).
\newblock Cold and trapped metastable noble gases.
\newblock {\em Rev. Mod. Phys.}, {\bf 84}, 175--210.

\bibitem[Viteau {\em et~al.}(2010)Viteau, Radogostowicz, Chotia, Bason,
  N.Malossi, Fuso, Ciampini, Morsch, Ryabtsev, and Arimondo]{Viteau2010}
Viteau, M., Radogostowicz, J., Chotia, A., Bason, M.~G., N.Malossi, Fuso, F.,
  Ciampini, D., Morsch, O., Ryabtsev, I.~I., and Arimondo, E. (2010).
\newblock Ion detection in the photoionization of a rb bose–einstein
  condensate.
\newblock {\em J. Phys. B\/}, {\bf 43}, 155301.

\bibitem[Vogler {\em et~al.}(2013)Vogler, Labouvie, Stubenrauch, Barontini,
  Guarrera, and Ott]{Vogler2013}
Vogler, A., Labouvie, R., Stubenrauch, F., Barontini, G., Guarrera, V., and
  Ott, H. (2013).
\newblock Thermodynamics of strongly correlated one-dimensional bose gases.
\newblock {\em Phys. Rev. A\/}, {\bf 88}, 031603.

\bibitem[Vuleti\'{c} {\em et~al.}(1998)Vuleti\'{c}, Chin, Kerman, and
  Chu]{Vuletic1998}
Vuleti\'{c}, V., Chin, C., Kerman, A.~J., and Chu, S. (1998).
\newblock Degenerate raman sideband cooling of trapped cesium atoms at very
  high atomic densities.
\newblock {\em Phys. Rev. Lett.}, {\bf 81}, 5768.

\bibitem[Weber {\em et~al.}(2006)Weber, Volz, Saucke, Kurtsiefer, and
  Weinfurter]{Weber2006}
Weber, M., Volz, J., Saucke, K., Kurtsiefer, C., and Weinfurter, H. (2006).
\newblock Analysis of a single-atom dipole trap.
\newblock {\em Phys. Rev. A\/}, {\bf 73}, 043406.

\bibitem[Weitenberg {\em et~al.}(2011)Weitenberg, Endres, Sherson, Cheneau,
  Schau\ss{}, Fukuhara, Bloch, and 1]{Weitenberg2011}
Weitenberg, C., Endres, M., Sherson, J.~F., Cheneau, M., Schau\ss{}, P.,
  Fukuhara, T., Bloch, I., and 1, S.~K. (2011).
\newblock Single-spin addressing in an atomic mott insulator.
\newblock {\em Nature\/}, {\bf 471}, 319.

\bibitem[Wilk {\em et~al.}(2010)Wilk, Gaetan, Evellin, Wolters, Miroshnychenko,
  Grangier, and Browaeys]{Wilk2010}
Wilk, T., Gaetan, A., Evellin, C., Wolters, J., Miroshnychenko, Y., Grangier,
  P., and Browaeys, A. (2010).
\newblock Entanglement of two individual neutral atoms using rydberg blockade.
\newblock {\em Phys. Rev. Lett.}, {\bf 104}, 010502.

\bibitem[Wineland {\em et~al.}(1973)Wineland, Ekstrom, and
  Dehmelt]{Wineland1973}
Wineland, D., Ekstrom, P., and Dehmelt, H. (1973).
\newblock Monoelectron oscillator.
\newblock {\em Phys. Rev. Lett.}, {\bf 31}, 1279.

\bibitem[Wineland and Itano(1981)Wineland and Itano]{Wineland1981}
Wineland, D.~J. and Itano, W.~M. (1981).
\newblock Spectroscopy of a single mg$^+$ ion.
\newblock {\em Phys. Lett.}, {\bf 82A}, 75.

\bibitem[Wiseman and Milburn(2010)Wiseman and Milburn]{Wiseman_2010}
Wiseman, H. and Milburn, G. (2010).
\newblock Quantum measurement and control.
\newblock {\em Cambridge University Press, Cambridge, England\/}.

\bibitem[Witthaut {\em et~al.}(2011)Witthaut, Trimborn, Hennig, Kordas, Geisel,
  and Wimberger]{Witthaut2011}
Witthaut, D., Trimborn, F., Hennig, H., Kordas, G., Geisel, T., and Wimberger,
  S. (2011).
\newblock Beyond mean-field dynamics in open bose-hubbard chains.
\newblock {\em Phys. Rev. A\/}, {\bf 83}, 063608.

\bibitem[W\"urtz {\em et~al.}(2009)W\"urtz, Langen, Gericke, Koglbauer, and
  Ott]{Wurtz_PRL_2009}
W\"urtz, P., Langen, T., Gericke, T., Koglbauer, A., and Ott, H. (2009).
\newblock Experimental demonstration of single-site addressability in a
  two-dimensional optical lattice.
\newblock {\em Phys. Rev. Lett.}, {\bf 103}, 080404.

\bibitem[W{\"urtz} {\em et~al.}(2010a)W{\"urtz}, Gericke, Vogler, Etzold, and
  Ott]{Wurtz_applPhysB2010}
W{\"urtz}, P., Gericke, T., Vogler, A., Etzold, F., and Ott, H. (2010a).
\newblock Image formation in scanning electron microscopy of ultracold atoms.
\newblock {\em Appl. Phys. B\/}, {\bf 98}, 641.

\bibitem[W{\"urtz} {\em et~al.}(2010b)W{\"urtz}, Gericke, Vogler, and
  Ott]{Wurtz_2010}
W{\"urtz}, P., Gericke, T., Vogler, A., and Ott, H. (2010b).
\newblock Ultracold atoms as a target: absolute scattering cross-section
  measurements.
\newblock {\em New. J. Phys.}, {\bf 12}, 065033.

\bibitem[Yamamoto {\em et~al.}(2016)Yamamoto, Kobayashi, Kuno, Kato, and
  Takahashi]{Yamamoto2016}
Yamamoto, R., Kobayashi, J., Kuno, T., Kato, K., and Takahashi, Y. (2016).
\newblock An ytterbium quantum gas microscope with narrow-line laser cooling.
\newblock {\em New J. Phys.}, {\bf 18}, 023016.

\bibitem[Yamazaki {\em et~al.}(2010)Yamazaki, Taie, Sugawa, and
  Takahashi]{Yamazaki2010}
Yamazaki, R., Taie, S., Sugawa, S., and Takahashi, Y. (2010).
\newblock Submicron spatial modulation of an interatomic interaction in a
  bose-einstein condensate.
\newblock {\em Phys. Rev. Lett.}, {\bf 105}, 050405.

\bibitem[Yan {\em et~al.}(2013)Yan, DeSalvo, Ramachandhran, Pu, and
  Killian]{Yan2013}
Yan, M., DeSalvo, B.~J., Ramachandhran, B., Pu, H., and Killian, T.~C. (2013).
\newblock Controlling condensate collapse and expansion with an optical
  feshbach resonance.
\newblock {\em Phys. Rev. Lett.}, {\bf 110}, 123201.

\bibitem[Ye {\em et~al.}(1999)Ye, Vernooy, and Kimble]{Ye1999}
Ye, J., Vernooy, D.~W., and Kimble, H.~J. (1999).
\newblock Trapping of single atoms in cavity qed.
\newblock {\em Phys. Rev. Lett.}, {\bf 83}, 4987.

\end{thebibliography}

\end{document}